\documentclass[11pt,a4paper]{article}

\usepackage{jheppub}						
\usepackage[T1]{fontenc}					
\usepackage{mathtools,physics} 				
\usepackage{subcaption} 			
\usepackage{booktabs,multirow}     
\usepackage{cleveref}						
\usepackage{orcidlink}						
\usepackage{bbm}

\crefname{figure}{figure}{figures}

\preprint{
  \begin{minipage}{5cm}
    \small
    \flushright
    KYUSHU-HET-359\\
    KANAZAWA-26-01
  \end{minipage}}

\title{Two component pseudo-Nambu-Goldstone-boson dark matter}

\author[a]{Riasat Sheikh\,\orcidlink{0009-0007-1207-1358},}
\author[b,c]{Takashi Toma\,\orcidlink{0000-0001-5828-0090},}
\author[a,d,e,f]{and Koji Tsumura\,\orcidlink{0000-0003-3765-2750}}

\affiliation[a]{Department of Physics, Kyushu University,\\ 744 Motooka, Nishi-ku, Fukuoka, 819-0395, Japan}
\affiliation[b]{Institute of Liberal Arts and Science, Kanazawa University,\\ Kanazawa 920-1192, Japan}
\affiliation[c]{Institute for Theoretical Physics, Kanazawa University,\\ Kanazawa 920-1192, Japan}
\affiliation[d]{Research Center for Advanced Particle Physics, Kyushu University, \\ 744 Motooka, Nishi-Ku, Fukuoka 819-0395, Japan}
\affiliation[e]{Quantum and Spacetime Research Institute, Kyushu University,\\ 744 Motooka, Nishi-Ku, Fukuoka 819-0395, Japan}
\affiliation[f]{Kavli IPMU (WPI), UTIAS, University of Tokyo,\\ Kashiwa, 277-8584, Japan}

\emailAdd{riasat.sheikh@phys.kyushu-u.ac.jp}
\emailAdd{toma@staff.kanazawa-u.ac.jp}
\emailAdd{tsumura.koji@phys.kyushu-u.ac.jp}

\date{\today}

\abstract{We study a two-component pseudo-Nambu-Goldstone-boson (pNGB) dark matter (DM) model motivated by boosted dark matter (BDM). The model is based on a complex scalar field charged under a dark $\text{U}(1)_V$ gauge symmetry, with a softly broken global $\text{SU}(3)_g$ symmetry that is spontaneously broken. The pNGB nature suppresses DM--Nucleon scattering, while the residual $\text{U}(1)_3 \times \text{U}(1)_{T_0}$ symmetry automatically stabilizes the two pNGB DM candidates and allows conversion of the heavier component into the lighter one. A central point is that the heavier or light component hierarchy is controlled by the two independent soft-breaking parameters that split the pNGB multiplet, so an abundant heavier component required for BDM can be obtained without introducing ad hoc hierarchies among independent portal coupling tuned to enable effective conversion. We analyze the relic abundance together with the constraints considered in this work, including Higgs invisible decays and perturbative unitarity, classify the coupled freeze-out dynamics, and assess the resulting BDM scattering cross section and flux.}


\begin{document}
\maketitle
\flushbottom

\section{Introduction}
\label{m2:sec:intro}

The existence of dark matter (DM) is one of the most compelling pieces of evidence for physics beyond the Standard Model (SM). The gravitational effects of DM are well-established, as evidenced by the rotation curves of galaxies and the large-scale structure of the universe. However, its particle nature remains elusive, with no direct detection of DM particles to date \cite{PandaX-4T:2021bab,LZ:2022lsv,XENON:2023cxc}. Among several viable candidates, the weakly interacting massive particle (WIMP) scenario has long served as a benchmark due to its natural consistency with thermal freeze-out production. However, null results from direct detection experiments have imposed increasingly stringent constraints on WIMP models, motivating the search for alternative mechanisms that naturally evade such limits.

A compelling class of candidates arises from pseudo-Nambu-Goldstone-boson (pNGB), which emerge from the spontaneous and explicit breaking of global symmetries \cite{Gross:2017dan}. Models for the pNGB DM are particularly attractive as their derivative-dominated interactions (in a non-linear representation of scalars) suppress scattering amplitudes at low momentum transfer, thereby remaining consistent with the latest bounds from direct detection experiments \cite{PandaX-4T:2021bab,LZ:2022lsv,XENON:2023cxc}. At the same time, these models retain sufficient annihilation cross section of the DM into SM particles  i.e., $\expval{\sigma v}_\text{ann.}^{} \simeq 10^{-26}~\text{cm}^3 \text{s}^{-1}$ to account for the observed relic density $\Omega_{\text{DM}} h^2 = 0.120 \pm 0.001$ \cite{Planck:2018vyg}.

The original pNGB model proposed in~\cite{Gross:2017dan} successfully addressed the direct detection problem but suffered from the domain wall (DW) issue (see also~\cite{Karamitros:2019ewv}). In the subsequent extended models~\cite{Abe:2020iph,Okada:2020zxo,Abe:2021byq,Okada:2021qmi,Abe:2022mlc,Liu:2022evb,Otsuka:2022zdy,Abe:2024vxz} (see also~\cite{Abe:2021nih,Abe:2021vat,Cai:2021evx,Cho:2023hek,Maji:2023fba}), the DW problem was avoided by embedding the softly broken symmetry into a gauge symmetry while preserving the key features of the pNGB framework. However, no detectable signal has been identified in these scenarios so far. More recently, \cite{Abe:2024lzj} investigated multi-component pNGB DM in a setup where, in addition to the pNGB, a dark gauge boson or a CP-odd scalar can act as a subdominant DM component. Although these extra components have unsuppressed WIMP--nucleon scattering, the effective direct-detection signal is strongly reduced by their small relic abundances, so the predicted event rates remain below current bounds.

Lying dormant for decades \cite{Silk:1985ax,Press:1985ug,Freese:1985qw,Krauss:1985aaa}, the idea of boosted dark matter (BDM) has recently gained traction as a promising avenue for exploring the nature of DM \cite{Toma:2021vlw,Miyagi:2022gvy,Aoki:2023tlb,BetancourtKamenetskaia:2025noa}. The underlying concept of BDM is that DM particles constituting the halo of the Milky Way galaxy can scatter with the nucleus of a massive celestial body, such as the Sun or the Earth, if their orbit passes through it. If their velocity after scattering is smaller than the escape velocity of the celestial body, they become gravitationally bound and start orbiting around it. Upon additional scattering they sink toward the center and accumulate, building up a local DM overdensity concentrated in a relatively small volume \cite{Baratella:2013fya}. There can then be various ways in which the DM particles collide with each other and produce highly energetic DM. In our previous work \cite{Sheikh:2020grh}, we explored the possibility of pNGB BDM via semi-annihilation. Here we instead focus on the conversion of a heavier DM component into a lighter one.

For BDM from a multi-component dark sector, one needs more than just a conversion channel: the heavier state must also survive today with an appreciable relic fraction. In generic multi-component models, the relic fractions are often controlled by several independent couplings and viable regions are frequently tied to ad hoc choices \cite{Bhattacharya:2013twoComp,Bian:2014multiHiggs,Nagao:2024hit}. It is therefore nontrivial to realize an abundant heavier component in a way that looks structural rather than tuned.

In this work, we propose a new pNGB DM model based on one SM-singlet complex scalar field which is a triplet under a global $\text{SU}(3)_g$ symmetry and charged under a dark $\text{U}(1)_V$ gauge symmetry. The spontaneous breaking of this symmetry, together with a soft breaking of the global symmetry, yields two pNGB states that play the role of DM candidates. The residual $\text{U}(1)_3^{} \times \text{U}(1)_{T_0}^{}$ symmetry automatically stabilizes the two DM candidates and allows annihilation of the heavier DM species into the lighter one, i.e., DM conversion. Their masses are controlled by the soft-breaking parameters: $m_8^2$ sets the common pNGB mass scale, while $m_3^2$ sets the splitting and determines which state is heavier. In this way, the mass hierarchy and the conversion channel responsible for BDM arise from the same symmetry structure, rather than from ad hoc tuning of independent portals. This process is forbidden in earlier pNGB models and gives rise to new phenomenological features for pNGB DM. One of the main motivations of this model is therefore to provide a concrete and natural pNGB realization of the BDM mechanism studied in \cite{Aoki:2023tlb,Toma:2021vlw,Miyagi:2022gvy,BetancourtKamenetskaia:2025noa}, and to assess how efficiently such a setup can generate a boosted flux.

This paper is organized as follows. In \cref{m2:sec:model}, we introduce the model and its Lagrangian, including the scalar potential and gauge kinetic mixing. We also discuss the mass spectrum and the parameters of the model. In \cref{m2:sec:constraints}, we analyze the constraints considered in this work, namely perturbative unitarity and Higgs invisible decay. In \cref{m2:sec:DM}, we study the conversion process that can lead to BDM signatures. We also explore the parameter space consistent with the relic abundance and the constraints imposed here, and estimate the resulting BDM flux. Finally, we conclude in \cref{m2:sec:conclusion}.


\section{The Model}
\label{m2:sec:model}

\subsection{Defining the Lagrangian}

We introduce a complex scalar field $S$ which is SM singlet and transforms under a gauged $\text{U}(1)_V$ symmetry as
\begin{equation}
  S \to e^{i \theta_V^{}(x)} S,
\end{equation}
where $\theta_V^{}(x)$ is the real-valued spacetime dependent gauge parameter and also transforms under a global $\text{SU}(3)_g$ symmetry with $\lambda_a$ being the generators of the group, as
\begin{equation}
  S \to e^{i \theta_a \lambda_a} S,
\end{equation}
The gauge and global charge assignments of the scalar fields $\Phi$ and $S$ are summarized in \cref{m2:tab:charges}. The Lagrangian of our model is given as
\begin{align}\label{m2:eq:7}
  \mathcal{L} & =
  \mathcal{L}_{\text{SM} \, \text{(w/o Higgs potential)}}
  + \abs{D_{\mu} S}^2
  - \frac{1}{4} V^{\mu \nu} V_{\mu \nu} - \frac{ \sin \epsilon}{2} V^{\mu \nu} B_{\mu \nu}
  - \mathcal{V}(S, \Phi),
\end{align}
where $\epsilon$ is the kinetic mixing angle and the covariant derivatives are defined as
\begin{align}\label{tcx-eq5}
  D_\mu S    & = \qty(\partial_\mu - i g_V^{} V_\mu)S,                                               \\
  D_\mu \Phi & = \qty(\partial_\mu - i \frac{g}{2} W^a_\mu \sigma^a - i \frac{g_Y^{}}{2} Y_\mu)\Phi,
\end{align}
where $\Phi$ is the SM Higgs doublet, and $V_\mu$ is the gauge field associated with the dark $\text{U}(1)_V$ gauge symmetry. $V_{\mu\nu}$ is the field strength tensor for the $\text{U}(1)_V$ symmetry gauge boson $V_\mu$. The scalar potential in \cref{m2:eq:7} of our model is given by
\begin{align}\label{m2:eq:8}
  \mathcal{V}(S, \Phi)
   & =
  \mu_S^2 \abs{S}^2
  +\frac{\lambda_S}{2} \abs{S}^4
  - \underbrace{\mu_\Phi^2 \abs{\Phi}^2
      + \frac{\lambda_\Phi}{2} \abs{\Phi}^4}_\text{SM Higgs}
  + \underbrace{\lambda_{\Phi S} \abs{\Phi}^2 \abs{S}^2}_\text{Higgs portal}
  + \underbrace{
      ~S^\dagger \, M_\text{soft}^2 \,S
    }_\text{soft breaking}\,.
\end{align}
Since the remaining scalar interactions depend only on $\abs{S}^2$, one may use an $\mathrm{SU}(3)_g$ basis transformation to diagonalize $M_\text{soft}^2$. Without any loss of generality, for the generic non-degenerate case relevant to two distinct pNGB states, the soft-breaking part may therefore be parameterized as
\begin{equation}
  \label{m2:eq:soft-breaking-diagonal}
  M_{\text{soft}}^2 = \frac{m_8^2}{\sqrt{3}} \lambda_8 + m_3^2 \lambda_3
  = \mathrm{diag}\qty(
  \frac{m_8^2}{3}+m_3^2,\,
  \frac{m_8^2}{3}-m_3^2,\,
  -\frac{2m_8^2}{3}),
\end{equation}
which explicitly breaks the global $\mathrm{SU}(3)_g$ symmetry down to $\mathrm{U}(1)_3 \times \mathrm{U}(1)_8$ and provides mass to the pNGBs as later seen in \cref{m2:sec:mass-spectrum}. We note that in this basis the scalar potential has a dark CP symmetry $S \to S^*$. Degeneracies occur for $m_3^2 = 0$ or $m_3^2 = \pm m_8^2$, in which case the residual symmetry is enhanced to $\mathrm{SU}(2)\times \mathrm{U}(1)$.

\begin{table}[t]
  \centering
  \begin{tabular}{lccccccc}
    \toprule
           & $\text{SU}(3)_c$ & $\text{SU}(2)_L$ & $\text{U}(1)_Y$ & $\text{SU}(3)_g$ & $\text{U}(1)_V$ \\
    \midrule
    $\Phi$ & $\mathbf{1}$     & $\mathbf{2}$     & $+\tfrac{1}{2}$ & $\mathbf{1}$     & $0$             \\
    $S$    & $\mathbf{1}$     & $\mathbf{1}$     & $0$             & $\mathbf{3}$     & $+1$            \\
    \bottomrule
  \end{tabular}
  \caption{Representation of $\Phi$ and $S$ complex scalar fields.}
  \label{m2:tab:charges}
\end{table}

\subsection{Residual symmetry in the broken phase}
\label{m2:sec:residual-symmetry}

Without any loss of generality, we can consider the vacuum expectation values (VEVs) for the singlet and the Higgs doublet as shown below (see \cref{m2:sec:vacuum_analysis})\footnote{Throughout this paper, matrices written with square brackets denote $\text{SU}(3)_g$ representations.}
\begin{equation}\label{m2:eq:2.6}
  \expval{S} = \frac{v_s^{}}{\sqrt{2}} \mqty[0 \\ 0 \\ 1],
  \quad
  \expval{\Phi} = \frac{v_{\Phi}^{}}{\sqrt{2}} \mqty(0 \\ 1),
\end{equation}
where $v_{\Phi}^{} = 1/\sqrt{\mathstrut \sqrt{2} G_F^{}}\approx246~\text{GeV}$ with the Fermi constant. This configuration spontaneously breaks the local gauge $\text{U}(1)_V$ symmetry. Therefore, we obtain the stationary conditions for $\mu_\Phi^2$ and $\mu_S^2$ as
\begin{align}
  \mu_\Phi^2
   & = \frac{1}{2} v_{\Phi}^2 \lambda_\Phi + \frac{1}{2} v_s^2 \lambda_{\Phi S}\label{m2:eq:14},              \\
  \mu_S^2
   & = \frac{2}{3} m_{8}^2 - \frac{1}{2} \qty(v_s^2 \lambda_S + v_{\Phi}^2 \lambda_{\Phi S})\label{m2:eq:15}.
\end{align}
Here, we realize that the soft-breaking term is engulfed by the $SU(3)_g$ invariant term, and the remaining soft breaking term is given by the diagonal generator $T_0$ as
\begin{equation}
  T_0 = \frac{2}{3} \mathbbm{1}_3 +  \frac{1}{\sqrt{3}} \lambda_8 = \text{diag}(1,1,0),
\end{equation}
which shows that, naturally after introducing the VEV of the complex scalar field $S$, we have $\text{U}(1)_3$ and $\text{U}(1)_{T_0}^{}$ residual symmetries ensuring the stability of the DM candidates. In this diagonal basis, the mass spectrum can be analyzed straightforwardly.

\subsection{Mass spectrum}
\label{m2:sec:mass-spectrum}

The SM Higgs doublet fluctuation can be defined as
\begin{equation}
  \Phi = \frac{1}{\sqrt{2}} \mqty(0 \\ v_{\Phi}^{} + h(x)),
\end{equation}
and for the complex triplet scalar field, only one of the component gains VEV and, without loss of generality, can be defined in the linear represenatation as,
\begin{equation}
  S = \mqty[S_1^{}\\ S_2^{}\\ S_3^{}] \to
  S = \frac{v_s^{}}{\sqrt{2}}\, \mqty[0 \\ 0 \\ 1] +
  \mqty[S_1^{}\\ S_2^{}\\ \cfrac{s(x) + iz(x)}{\sqrt{2}} ],
\end{equation}
where $z$ is the would-be Nambu-Goldstone (NG) boson absorbed by the $\text{U}(1)_V$ gauge boson.This allows a straightforward diagonalization of the mass matrix in the $\text{U}(1)_{3} \times \text{U}(1)_{T_0}^{}$ uncharged sector, yielding the physical mass eigenstates $h_1$ and $h_2$ as
\begin{equation}
  \mqty(m_{h_1}^2 & 0 \\ 0 & m_{h_2}^2) =
  \mqty(
  \cos\theta &  \sin\theta\\
  -  \sin\theta &  \cos\theta
  )\mqty(
  v_\Phi^2 \, \lambda_\Phi  &  v_s^{}v_\Phi^{}\lambda_{\Phi S}\\
  v_s^{}v_\Phi^{}\lambda_{\Phi S} & v_s^2 \lambda_S
  )\mqty(
  \cos\theta & -  \sin\theta\\
  \sin\theta &  \cos\theta
  ),
\end{equation}
and the mass eigenstate basis can be represented as
\begin{equation}
  \mqty( h_1^{}\\ h_2) =   \mqty(
  \cos\theta &  \sin\theta\\
  -  \sin\theta &  \cos\theta
  )\mqty(h \\ s).
\end{equation}
Furthermore, the mixing angle $\theta$ for the mass eigenstates $h_1$ and $h_2$ is given by
\begin{equation}
  \tan 2\theta = \frac{2 v_s^{} v_{\Phi}^{}\lambda_{\Phi S}}{v_{\Phi}^2 \, \lambda_\Phi -  v_s^2 \lambda_S} ~.
\end{equation}

On the other hand, the fields $S_1^{}$ and $S_2^{}$ carry charges $(+1,+1)$ and $(-1,+1)$, respectively, under the residual $\text{U}(1)_3 \times \text{U}(1)_{T_0}$ symmetry. These residual charges stabilize the DM candidates. Therefore, the mass matrix in terms of the mass eigenstates can be written as
\begin{align}
  m_{S_1}^2    & = m_8^2 + m_3^2 \equiv m_{\text{DM}_h^{}}^{2},     \\
  m_{S_2^{}}^2 & = m_8^2 - m_3^2 \equiv m_{\text{DM}_\ell^{}}^{2} .
\end{align}

Thus, this setup yields two complex pNGBs, $S_1^{}$ and $S_2^{}$. These fields do not mix with the neutral scalar sector and therefore provide natural two-component DM species that interact weakly with visible matter through Higgs portal interactions. Their masses are controlled by the soft breaking parameters $m_8^2$ and $m_3^2$, where $m_8^2$ sets the common pNGB mass scale and $m_3^2$ sets the mass splitting. The sign of $m_3^2$ only determines which state is heavier (see, \cref{m2:sec:vacuum_analysis}), so without loss of generality we take $m_3^2>0$ and identify $S_1^{} ~(\equiv S_h^{})$ as the heavier state and $S_2^{} ~(\equiv S_\ell^{})$ as the lighter one.

\begin{figure}[t]
  \centering
  \includegraphics[width=0.8\textwidth]{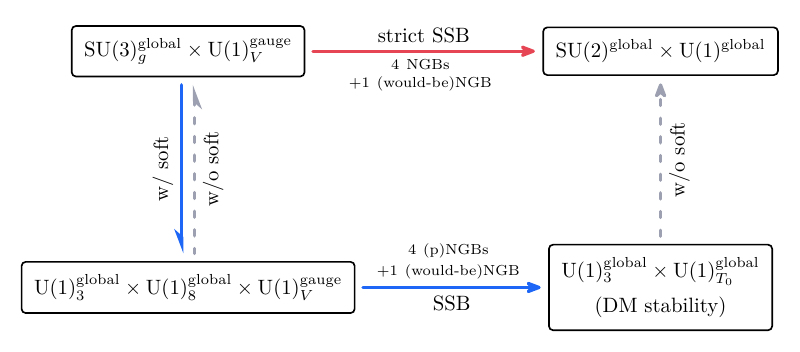}
  \caption{Summary of the symmetry-breaking pattern of our model. }
  \label{m2:fig:symmetry-summary-main}
\end{figure}

Therefore, the mass terms in the Lagrangian \cref{m2:eq:7} takes the form
\begin{equation}
  \mathcal{L} \supset -\frac{1}{2}m_{h_1}^2 h_1^2 -\frac{1}{2}m_{h_2}^2 h_2^2 - m_{\text{DM}_h^{}}^2 \abs{S_h^{}}^2 - m_{\text{DM}_\ell^{}}^2 \abs{S_\ell^{}}^2.
\end{equation}
In \cref{m2:fig:symmetry-summary-main}, we summarize the symmetry-breaking pattern of our model. Along the strict SSB path, the symmetry $\mathrm{SU}(3)_g^{\rm global}\times \mathrm{U}(1)_V^{\rm gauge}$ is spontaneously broken to $\mathrm{SU}(2)^{\rm global}\times \mathrm{U}(1)^{\rm global}$, yielding four physical NGBs and one would-be NGB. In the presence of soft-breaking terms, the symmetry is first explicitly reduced to $\mathrm{U}(1)_3^{\rm global}\times \mathrm{U}(1)_8^{\rm global}\times \mathrm{U}(1)_V^{\rm gauge}$, and the subsequent SSB leaves an unbroken $\mathrm{U}(1)_3^{\rm global}\times \mathrm{U}(1)_{T_0}^{\rm global}$ subgroup. In this case, the four NGBs become pNGBs, while the would-be NGB is associated with the broken gauge direction, and the residual $\mathrm{U}(1)_3^{\rm global}\times \mathrm{U}(1)_{T_0}^{\rm global}$ ensures DM stability.

\subsection{Parameters of our model}

There are total nine parameters i.e., seven from the scalar sector and two from the new dark gauge sector in this model.
Apart from the stationary conditions in \cref{m2:eq:14} and \cref{m2:eq:15}, here we will write all the parameters of our model in terms of the physical mas  `s eigenvalues, mixing angle and VEVs as
\begin{subequations}
  \begin{align}
    m_8^2  & = \frac{1}{2} \qty(m_{\text{DM}_h^{}}^{2} + m_{\text{DM}_\ell^{}}^{2}),                                            \\
    m_3^2  & = \frac{1}{2} \qty(m_{\text{DM}_h^{}}^{2} - m_{\text{DM}_\ell^{}}^{2}),                                            \\
    \lambda_\Phi
           & = \frac{m_{h_1}^2  \cos^2\theta + m_{h_2}^2  \sin^2\theta}{v_{\Phi}^{2}}\label{m2:eq:2.51},                        \\
    \lambda_{\Phi S}
           & {}= \qty(\cfrac{v_{\Phi}^{}}{v_s^{}})^2
    \qty(\frac{m_{h_1}^2 - m_{h_2}^2 }{v_{\Phi}^{2}})
    \sin\theta  \cos\theta,                                                                                                     \\
    \lambda_S
           & =   \qty(\cfrac{v_{\Phi}^{}}{v_s^{}})^2\frac{m_{h_1}^2  \sin^2\theta + m_{h_2}^2  \cos^2\theta}{v_{\Phi}^{2}},     \\
    g_V^{} & \simeq \qty(\cfrac{v_{\Phi}^{}}{v_s^{}})^{} \frac{m_{Z'}^{}}{v_{\Phi}^{}}, \quad \text{with} \quad \epsilon \ll 1.
    \label{m2:eq:2.54}
  \end{align}
\end{subequations}
Here, $g_V^{}$ is the dark gauge coupling and $\epsilon$ is the gauge kinetic mixing angle (see, \cite{Abe:2024vxz}). Thus, we are left with the following physical parameters
\begin{equation}
  m_{h_1}^{}( = 125 \;\text{GeV}),\quad m_{h_2}^{},\quad \sin\theta,\quad m_{\text{DM}_h^{}}^{},\quad
  m_{\text{DM}_\ell^{}}^{},\quad v_{\Phi}^{},
  \quad v_s^{}, \quad m_{Z'}^{}, \quad \sin\epsilon.
  \label{m2:eq:free-parameters}
\end{equation}
We can therefore perform a parameter scan over the dimensionless ratio $\flatfrac{v_{\Phi}^{}}{v_s^{}}$ to understand various properties of the DM interactions. From here onwards, we will refer to the coupling parameters in terms of the aforementioned dimensionless quantity.


\section{Constraints on the model}
\label{m2:sec:constraints}

In this section, we discuss the constraints on the model parameters that are most relevant for the present BDM analysis, namely perturbative unitarity (PU) and the Higgs invisible decay. A dedicated direct-detection analysis is beyond the scope of this work, since the pNGB nature of the DM suppresses the DM--nucleon scattering amplitude at low momentum transfer, as established in the pNGB literature.

\subsection{Perturbative unitarity}
\label{m2:sec:constraints.1}
The PU of the model is a crucial aspect to ensure the stability of the theory. From the discussion of PU \cite{Lee:1977eg}, we can obtain the constraints on the model parameters. Since we are going to deal with high energy scattering process, we study the Lagrangian in the symmetric phase, i.e., from \cref{m2:eq:8},
\begin{equation}
  \mathcal{V} \supset \frac{\lambda_\Phi}{2}\,(\Phi^\dagger \Phi)^2
  + \frac{\lambda_S}{2}\,(S^\dagger S)^2
  + \lambda_{\Phi S}\,(\Phi^\dagger \Phi)(S^\dagger S).
\end{equation}
For this purpose, we define the SM Higgs doublet and the complex scalar triplet
as,
\begin{equation}
  \Phi = \mqty(\phi_1 \\ \phi_2),
  \quad
  S = \mqty[S_1^{}\\ S_2^{}\\ S_3^{}],
\end{equation}
and the above potential can be re-written as
\begin{align}
  \mathcal{V} \supset
   & \frac{\lambda_\Phi}{2} \qty(\phi_1 \phi_1^* + \phi_2 \phi_2^*)^2
  + \frac{\lambda_S}{2} \qty(S_1^{}S_1^* + S_2^{}S_2^* + S_3^{}S_3^*)^2
  \nonumber                                                                                                    \\
   & + \lambda_{\Phi S} \qty(\phi_1 \phi_1^* + \phi_2 \phi_2^*) \qty(S_1^{}S_1^* + S_2^{}S_2^* + S_3^{}S_3^*).
\end{align}
Therefore, we can construct the charge-neutral states under the $\text{U}(1)_V$ gauge symmetry as
\begin{equation}
  i\to f \quad \forall \quad i,f \in \qty{\phi_1 \phi_1^*,\, \phi_2 \phi_2^*,\, S_1^{}S_1^*,\, S_2^{}S_2^*,\, S_3^{}S_3^*,\, S_1^{}S_2^*,\, S_2^{}S_3^*,\, S_3^{}S_1^* }.
\end{equation}
The partial wave matrix $a^0$ thus can be constructed as
\begin{align}
  (a^0)_{fi} = \frac{1}{16 \pi} \mqty(
  2\lambda_\Phi    & \lambda_\Phi     & \lambda_{\Phi S} & \lambda_{\Phi S} & \lambda_{\Phi S} & 0         & 0         & 0         \\
  \lambda_\Phi     & 2\lambda_\Phi    & \lambda_{\Phi S} & \lambda_{\Phi S} & \lambda_{\Phi S} & 0         & 0         & 0         \\
  \lambda_{\Phi S} & \lambda_{\Phi S} & 2\lambda_S       & \lambda_S        & \lambda_S        & 0         & 0         & 0         \\
  \lambda_{\Phi S} & \lambda_{\Phi S} & \lambda_S        & 2\lambda_S       & \lambda_S        & 0         & 0         & 0         \\
  \lambda_{\Phi S} & \lambda_{\Phi S} & \lambda_S        & \lambda_S        & 2\lambda_S       & 0         & 0         & 0         \\
  0                & 0                & 0                & 0                & 0                & \lambda_S & 0         & 0         \\
  0                & 0                & 0                & 0                & 0                & 0         & \lambda_S & 0         \\
  0                & 0                & 0                & 0                & 0                & 0         & 0         & \lambda_S
  ).
\end{align}
Calculating the eigenvalues of this matrix and imposing the PU conditions we find the following inequalities
\begin{subequations}
  \begin{align}
    \abs{\lambda_\Phi}                       & < 8\pi,  \\
    \abs{\lambda_S}                          & < 8 \pi, \\
    \abs{\lambda_{\Phi S}}                   & < 8 \pi, \\
    \abs{4 \lambda_S + 3 \lambda_\Phi
      \pm \sqrt{
        \qty(4 \lambda_S - 3 \lambda_\Phi)^2
        + 24 \lambda_{\Phi S}^2
      }
    } & < 16 \pi.
  \end{align}
\end{subequations}
From the charged states under the $\text{U}(1)_V$ symmetry, we have the channels such as $S_1^{}S_1^{}\leftrightarrow S_1^{}S_1^{}$ and $S_1^{}\phi_1^{} \leftrightarrow S_1^{}\phi_1^{}$, which will give us the following inequalities:
\begin{subequations}
  \begin{align}
    \abs{\lambda_S}        & < 8 \pi, \\
    \abs{\lambda_{\Phi S}} & < 8 \pi.
  \end{align}
\end{subequations}
The PU bounds on the dark gauge coupling is similar to the ones in \cite{Abe:2024vxz}, i.e.,
\begin{equation}
  g_V^{} < \sqrt{4 \pi}.
\end{equation}

\subsection{Higgs invisible decay}

\begin{figure}[t]
  \centering
  \begin{equation*}
    \raisebox{-0.45\height}{\includegraphics[width=2.6cm]{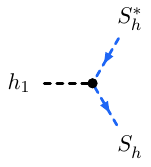}}
    = \kappa_{1}^{}
    \qquad\quad
    \raisebox{-0.45\height}{\includegraphics[width=2.6cm]{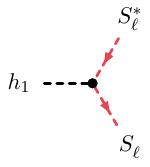}}
    = \kappa_{1}^{}
  \end{equation*}

  \caption{SM Higgs--DM--DM vertex}
  \label{m2:fig:DM-Higgs-vertex}
\end{figure}

A stringent constraint in the low-mass region of the parameter space arises from the Higgs invisible decay width, particularly when $m_{\mathrm{DM_j}^{}} < \flatfrac{m_{h_1}^{}}{2}$, where $j = h,\ell$. In this case, the Higgs boson $h_1$ can decay into a pair of either of the DM species contributing to the Higgs invisible decay channel. The corresponding vertex is shown in \cref{m2:fig:DM-Higgs-vertex} and can be written as
\begin{equation}
  \label{m2:eq:DM-Higgs-vertex}
  \kappa_{1}^{}
  = v_{\Phi}^{} \lambda_{\Phi S} \cos \theta  + v_s^{} \lambda_S \sin \theta
  = \qty(\cfrac{v_{\Phi}^{}}{v_s^{}})\cfrac{m_{h_1}^2}{v_{\Phi}^{}}\sin\theta.
\end{equation}
Thus the decay width for each DM species can be calculated as
\begin{align}
  \Gamma_{h_1^{}\to S_j^* S_j^{}}^{}
  ={} & \qty(\cfrac{v_{\Phi}^{}}{v_s^{}})^{2}
  ~\frac{\sin^2\theta \,m_{h_1}^{3}}{16 \,\pi \,v_{\Phi}^{2}}
  \sqrt{
    1 - \cfrac{4 m_{\text{DM}_j^{}}^2}{m_{h_1}^{2}}
  }
  ~\Theta\qty(m_{h_1}^{} - 2 m_{\text{DM}_j^{}}^{}),
\end{align}
where, $j = h, \ell$ and $\Theta(m_{h_1}^{} - 2 m_{\text{DM}_j^{}}^{})$ is the Heaviside step function, which ensures that the decay only occurs when $m_{h_1}^{} > 2 m_{\text{DM}_j^{}}^{}$.
The total invisible decay width into the DM particles is given by
\begin{equation}
  \Gamma_{h_1}^{\text{inv}}
  = \sum_j  \Gamma_{h_1^{}\to S_j^* S_j^{}}^{}, \quad \text{with} \quad j = h, \ell.
\end{equation}
Therefore, the total decay width of $h_1$ becomes
\begin{equation}
  \Gamma_{h_1}^{\text{tot}}
  = \cos^2\theta ~\Gamma_{h_1}^{\text{SM}} + \Gamma_{h_1}^{\text{inv}},
\end{equation}
where $\Gamma_{h_1}^{\text{SM}} \approx 4.07~\text{MeV}$ is the Higgs--SM  total decay width at $m_{h_1}=125~\text{GeV}$. This process is being searched by ATLAS and CMS experiments with the upper bound currently at,
\begin{equation}
  \text{BR}_\text{inv}^{}
  \equiv \cfrac{
    \Gamma_{h_1}^{\text{inv}}
  }{\Gamma_{h_1}^{\text{tot}}}
  <
  \begin{cases}
    0.107 & (\text{ATLAS \cite{ATLAS:2023tkt}}) \\
    0.15  & (\text{CMS \cite{CMS:2023sdw}})
  \end{cases}.
\end{equation}


\section{Dark matter}
\label{m2:sec:DM}

\subsection{New channels}
\label{m2:sec:new_channels}

\begin{figure}[t]
  \centering
  \raisebox{0.12\height}{\includegraphics[width=3.5cm]{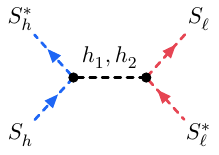}}
  \qquad
  \raisebox{0.12\height}{\includegraphics[width=3.5cm]{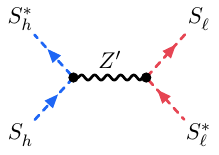}}
  \qquad
  \includegraphics[width=3.2cm]{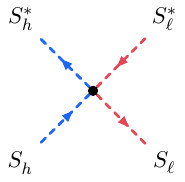}
  \caption{DM--DM conversion processes.}
  \label{m2:fig:DM-conversion}
\end{figure}

In this two-component pNGB DM model, the dark sector contains two stable DM species, $S_h^{}$ and $S_\ell^{}$. The relevant interactions are given by
\begin{align}
  \label{m2:eq:new_channels_lagrangian}
  \mathcal{L} \supset
  \kappa_3^{} \qty\Big(S_h^* \overset{\leftrightarrow}{\partial_\mu} S_h^{}
  + S_\ell^* \overset{\leftrightarrow}{\partial_\mu} S_\ell^{}) Z'^\mu
  -  \qty\Big(\abs{S_h}^2 + \abs{S_\ell^{}}^2) \qty(\kappa_1^{} h_1^{}+ \kappa_2^{} h_2)
  - \kappa_4^{} \abs{S_h}^2 \abs{S_\ell^{}}^2,
\end{align}
where $A\overset{\leftrightarrow}{\partial_\mu}B=A\partial_\mu B-B\partial_\mu A$, and the coupling $\kappa_1^{}$ is as in \cref{m2:eq:DM-Higgs-vertex}, while the others are given by
\begin{align}
  \kappa_2^{} ={}       \qty(\cfrac{v_{\Phi}^{}}{v_s^{}})^{}\frac{m_{h_2}^2}{v_{\Phi}^{}} \cos\theta,
  \quad
  \kappa_3^{} \simeq{} \qty(\cfrac{v_{\Phi}^{}}{v_s^{}})^{}\cfrac{m_{Z'}^{}}{v_{\Phi}^{}},
  \quad
  \kappa_4^{} ={} \qty(\cfrac{v_{\Phi}^{}}{v_s^{}})^{2}\frac{m_{h_1}^2  \sin^2 \theta + m_{h_2}^2  \cos^2 \theta}{v_{\Phi}^{2}}.
  \label{m2:eq:kappas}
\end{align}
These couplings encapsulate the DM--DM interactions that are unique to our model, which induces the conversion processes, i.e.,
\begin{equation}
  \label{m2:eq:conversion_processes}
  S_h^* S_h^{}\to \qty{h_1,h_2,\, Z',\, \text{contact}} \to S_\ell^* S_\ell^{},
\end{equation}
whose Feynman diagrams are shown in \cref{m2:fig:DM-conversion}.

As the DM species share a common set of couplings, the interactions in \cref{m2:eq:kappas} play a crucial role during freeze-out in determining the relic abundance of each species through both annihilation and conversion processes. Consequently, the system exhibits coupled Boltzmann dynamics. Depending on the relative sizes of annihilation and conversion rates, the dark sector can exhibit collective freeze-out of the total abundance, sequential freeze-out with a late-time conversion source, or approximately independent freeze-out of two species.

These interactions provide an alternative mechanism for generating boosted pNGB DM and can lead to qualitatively distinct indirect detection signatures. They also allow viable relic density even where standard annihilation channels are inefficient.

\subsection{Relic abundance}
\label{m2:sec:relic}

The relic abundance of the two DM species $S_h^{}$ and $S_\ell^{}$ can be obtained by solving the Boltzmann equation including the following processes:
\begin{enumerate}
  \item As shown in \cref{m2:fig:DM-conversion}, the heavier DM ($S_h^{}$) can convert into the lighter one ($S_\ell^{}$) via Higgs bosons, dark gauge boson mediators, and a contact interaction.
  \item Moreover, both DM species can annihilate into pairs of SM particles, including the dark gauge boson, which is kinetically mixed with the SM hypercharge gauge boson.
\end{enumerate}
Analytically, these equations can be written schematically as
\begin{subequations}
  \label{m2:eq:boltzmann-main}
  \begin{align}
    \dv{Y_h^{}}{z}
     & =
    - \overline{\expval{\sigma v}}_{S_h^* S_h^{} \to \text{SM} \,\text{SM}}^{} \qty(Y_h^2-Y_{h,\text{eq}}^2)
    - \overline{\expval{\sigma v}}_{S_h^* S_h^{} \to S_\ell^* S_\ell^{}}^{} \qty{Y_h^2-\qty(\frac{Y_{h,\text{eq}}}{Y_{\ell,\text{eq}}})^2Y_\ell^2},
    \\
    \dv{Y_\ell^{}}{z}
     & =
    - \overline{\expval{\sigma v}}_{S_\ell^* S_\ell^{} \to \text{SM} \,\text{SM}}^{} \qty(Y_\ell^2-Y_{\ell,\text{eq}}^2)
    + \overline{\expval{\sigma v}}_{S_h^* S_h^{} \to S_\ell^* S_\ell^{}}^{} \qty{Y_h^2-\qty(\frac{Y_{h,\text{eq}}}{Y_{\ell,\text{eq}}})^2Y_\ell^2},
  \end{align}
\end{subequations}
where, the comoving number density $Y_{j (=\,h,\,\ell)}^{} = \flatfrac{n_j^{}}{\mathfrak{s}(T)}$, $z = \flatfrac{m_{\text{DM}_\ell^{}}}{T}$ and introduce the shorthand notation
\begin{equation}
  \label{m2:eq:cross-section-shorthand}
  \overline{\expval{\sigma v}}_{ijkl}^{}
  = \cfrac{\mathfrak{s}(T)}{H(T) ~z}
  \expval{\sigma v}_{ijkl}^{} \,,
\end{equation}
where $\mathfrak{s}(T)$ and $H(T)$ are the entropy density and the Hubble rate, respectively. The conversion process redistributes the abundance between the two DM species without changing the total dark number, as reflected in the second term of the Boltzmann equations. In \cref{m2:eq:cross-section-shorthand}, the indices $(ijkl)$ label the processes $ij \to kl$.

\begin{table}[t]
  \centering
  \begin{tabular}{lccccccclcc}
    \toprule
                         & $m_{\text{DM}_h^{}}^{}$ & $m_{\text{DM}_\ell^{}}^{}$ & $\gamma = \cfrac{m_{\text{DM}_h^{}}^{}}{m_{\text{DM}_\ell^{}}^{}}$ & $v_s^{}$                 & $\xi_j^{} = \cfrac{\Omega_j^{}}{\Omega_{\text{tot}}^{} }$               & Channel & $\ev{\sigma v}_{\text{ann.}}^{}$ \\
                         & [GeV]                   & [GeV]                      &                                                                    & [GeV]                    & $(j = h , \ell)$                                                        &         & [$\text{cm}^3\text{s}^{-1}$]     \\
    \midrule
    \multirow{3}{*}{RP1} & \multirow{3}{*}{125.62} & \multirow{3}{*}{100.50}    & \multirow{3}{*}{1.25}                                              & \multirow{3}{*}{501.067} & \multirow{3}{*}{\shortstack{$\xi_h^{} = 0.013$\\$\xi_\ell^{} = 0.987$}} & A       & $1.15\times 10^{-25}$            \\
                         &                         &                            &                                                                    &                          &                                                                         & B       & $2.02\times 10^{-26}$            \\
                         &                         &                            &                                                                    &                          &                                                                         & C       & $1.53\times 10^{-24}$            \\
    \midrule
    \multirow{3}{*}{RP2} & \multirow{3}{*}{5000}   & \multirow{3}{*}{500}       & \multirow{3}{*}{10}                                                & \multirow{3}{*}{218.408} & \multirow{3}{*}{\shortstack{$\xi_h^{} = 0.981$\\$\xi_\ell^{} = 0.019$}} & A       & $1.47\times 10^{-26}$            \\
                         &                         &                            &                                                                    &                          &                                                                         & B       & $1.42\times 10^{-24}$            \\
                         &                         &                            &                                                                    &                          &                                                                         & C       & $8.10\times 10^{-27}$            \\
    \bottomrule
  \end{tabular}
  \caption{Representative points for the dark matter yield evolution. The channel A, B and C correspond to $S_h^* S_h^{} \to \text{SM} \,\text{SM}$, $S_\ell^* S_\ell^{} \to \text{SM} \,\text{SM}$ and $S_h^* S_h^{} \to S_\ell^* S_\ell^{}$, respectively.}
  \label{m2:tab:benchmarks}
\end{table}

We use \texttt{FeynRules} \cite{Alloul:2013bka} to generate the model files and \texttt{micrOMEGAs} \cite{Alguero:2023zol} to compute the relic abundance numerically by solving the exact coupled system in \cref{m2:eq:boltzmann-main} for the DM yields. To illustrate the interplay between annihilation and conversion, we introduce representative points (RPs) in \cref{m2:tab:benchmarks}, where the mass ratio $\gamma = \flatfrac{m_{\text{DM}_h^{}}^{}}{m_{\text{DM}_\ell^{}}^{}}$ characterises the boost factor of the lighter DM species, used later in \cref{m2:sec:indirect_detection}. The relic fraction $\xi_j^{} = \flatfrac{\Omega_j^{}}{\Omega_{\text{tot}}^{}}$ ($j = h, \ell$) for each DM species is also evaluated at present. Since the same couplings in \cref{m2:eq:kappas} govern both processes for the two species, the corresponding rates are correlated through the common ratio $v_\Phi/v_s$, and the relative heavy and light DM composition is determined by the coupled Boltzmann system once the observed relic density is fixed. Additionally, for these points, we fix the remaining free parameters (see \cref{m2:eq:free-parameters}) as $\sin\theta = 0.1$, $m_{h_2} = 300~\text{GeV}$, $m_{Z'} = 200~\text{GeV}$, and $\sin\epsilon = 10^{-4}$, while the dark VEV $v_s$ is chosen to reproduce the observed relic abundance, $\Omega h^2 = 0.120 \pm 0.001$.

\begin{figure}[t]
  \centering
  \begin{subfigure}{0.49\textwidth}
    \includegraphics[width=\linewidth]{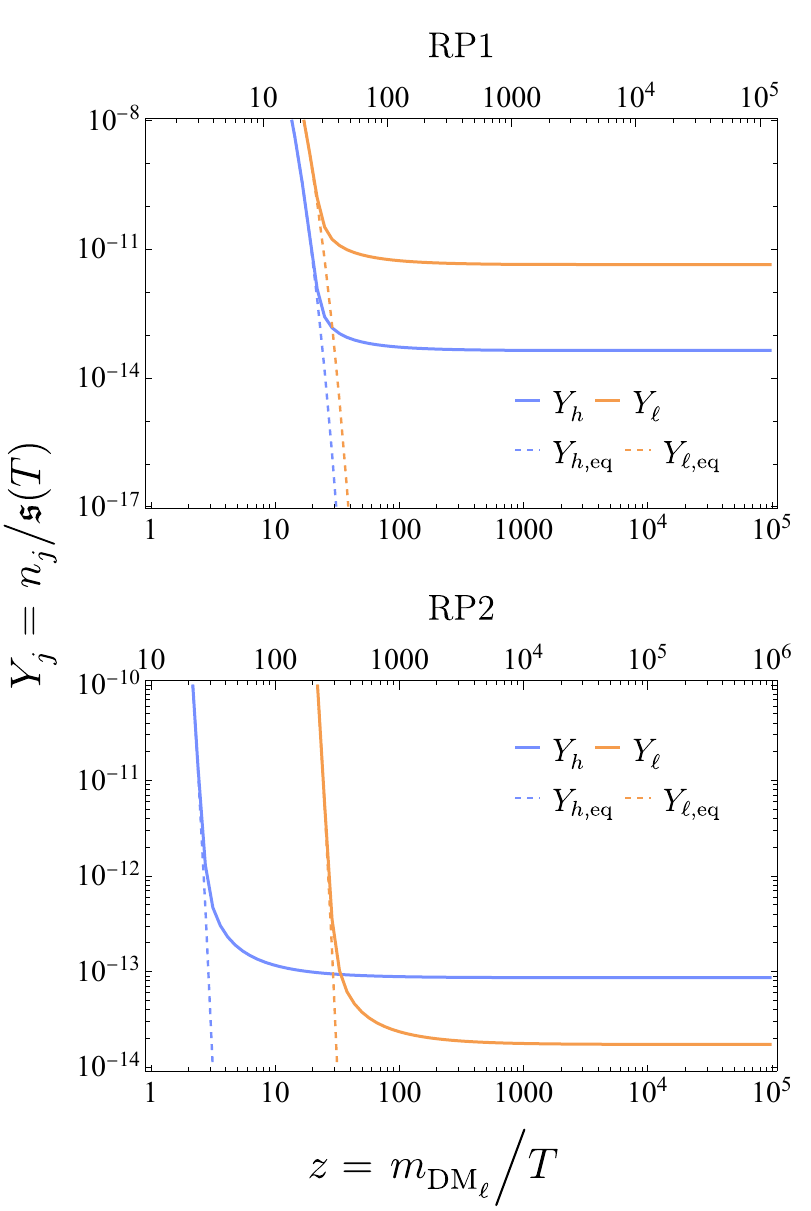}
    \caption{DM species yields.}
    \label{m2:fig:yeildvstemp}
  \end{subfigure}
  \begin{subfigure}{0.49\textwidth}
    \includegraphics[width=\linewidth]{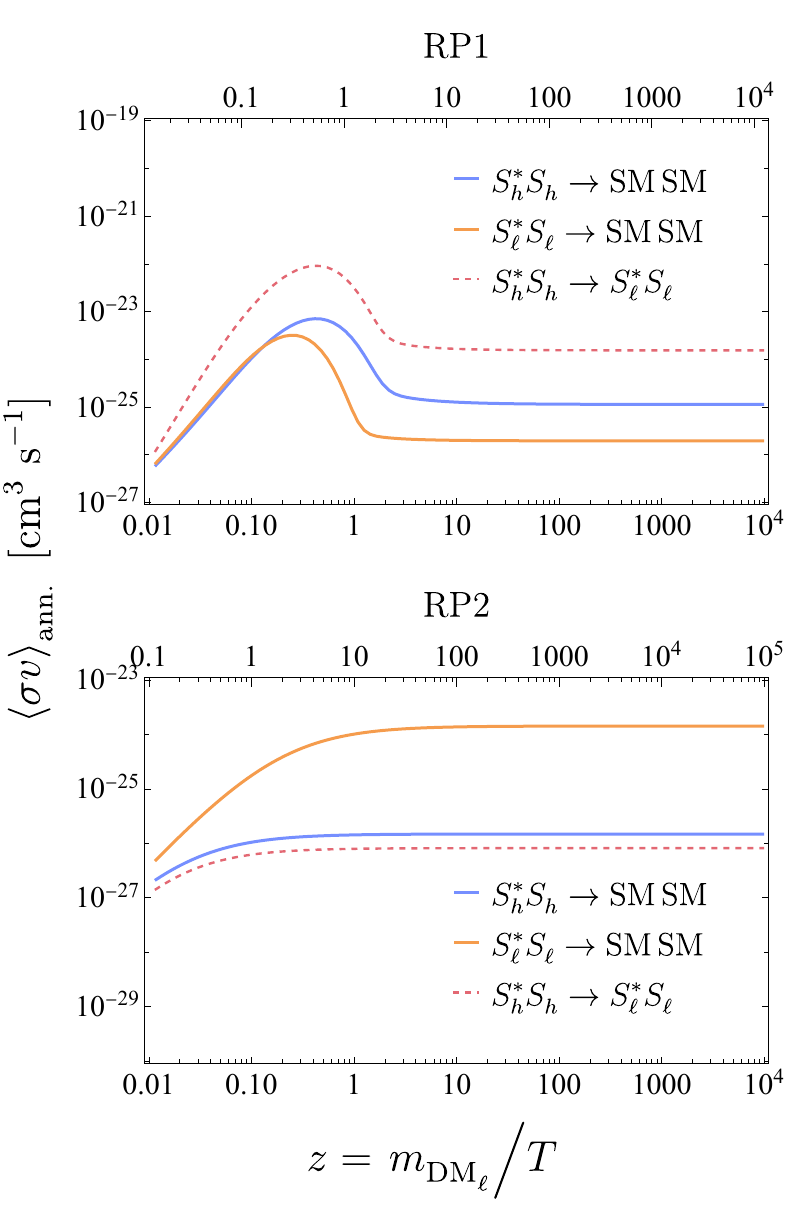}
    \caption{Annihilation and conversion cross sections.}
    \label{m2:fig:csvstemp}
  \end{subfigure}
  \caption{Temperature evolution of the dark matter yields and relevant cross sections for the representative points in \cref{m2:tab:benchmarks}. The top axis is rescaled as $m_{\text{DM}_h^{}}^{}/T = \gamma\, m_{\text{DM}_\ell^{}}^{}/T$. The index $j = h, ~\ell$ labels the two DM species.}
\end{figure}

The corresponding numerical yields and relevant cross sections for the RPs are shown in \cref{m2:fig:yeildvstemp,m2:fig:csvstemp}. In panel~(\subref{m2:fig:yeildvstemp}), solid and dashed lines denote the numerical and equilibrium yields of the heavier (blue) and lighter (orange) species, respectively. In panel~(\subref{m2:fig:csvstemp}), the blue and orange solid lines denote annihilation of the heavier and lighter DM species, respectively, while the red dashed line denotes the conversion process.

From \cref{m2:tab:benchmarks}, it is clear that the conversion process remains non-negligible. For a smaller boost factor, as in RP1 with $\gamma = 1.25$, corresponding to $m_{\text{DM}_h^{}}^{} = 125.62~\text{GeV}$ and $m_{\text{DM}_\ell^{}}^{} = 100.50~\text{GeV}$, the thermally averaged cross sections are enhanced in the small-$x$ region (\cref{m2:fig:csvstemp}), as thermal averaging samples a nearby resonant region. However, both species still closely track their equilibrium yields at that stage, and no visible replenishment of the light-species yield appears in RP1 (\cref{m2:fig:yeildvstemp}). The final heavy ($\xi_h^{} = 0.013$) to light ($\xi_\ell^{} = 0.987$) abundance ratio is determined later, around freeze-out, while conversion remains efficient until it eventually decouples.

For a larger boost factor, as in RP2 with $\gamma = 10$, corresponding to $m_{\text{DM}_h^{}}^{} = 5000~\text{GeV}$ and $m_{\text{DM}_\ell^{}}^{} = 500~\text{GeV}$, the light-species annihilation channel (\cref{m2:fig:csvstemp}) becomes increasingly important relative to the conversion process. Although the conversion rate is reduced, it remains non-negligible. Consequently, once the heavier component departs from equilibrium, the lighter species undergoes more efficient depletion (\cref{m2:fig:yeildvstemp}), resulting in an abundance ratio of $\xi_h^{} = 0.981$ and $\xi_\ell^{} = 0.019$. Far from the representative-point regions, the model can also approach a weak-conversion limit if the dark gauge boson channels are turned off; however, this regime is not central to the present BDM discussion.

\begin{figure}[t]
  \centering
  \begin{subfigure}{0.5\textwidth}
    \includegraphics[width=\linewidth]{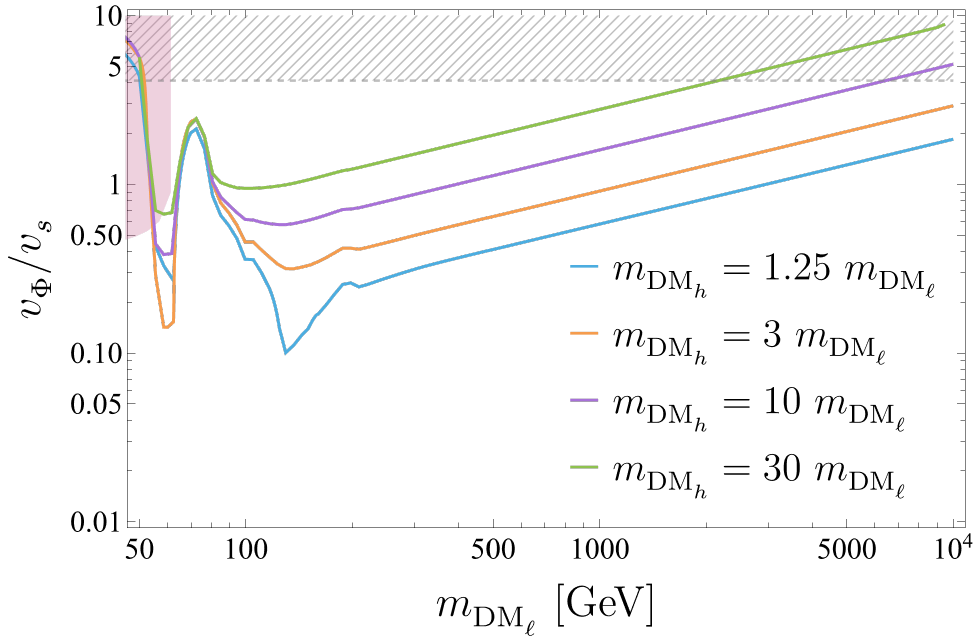}
    \caption{Light DM species mass vs. coupling.}
    \label{m2:fig:relic abundance1}
  \end{subfigure}
  \begin{subfigure}{0.5\textwidth}
    \includegraphics[width=\linewidth]{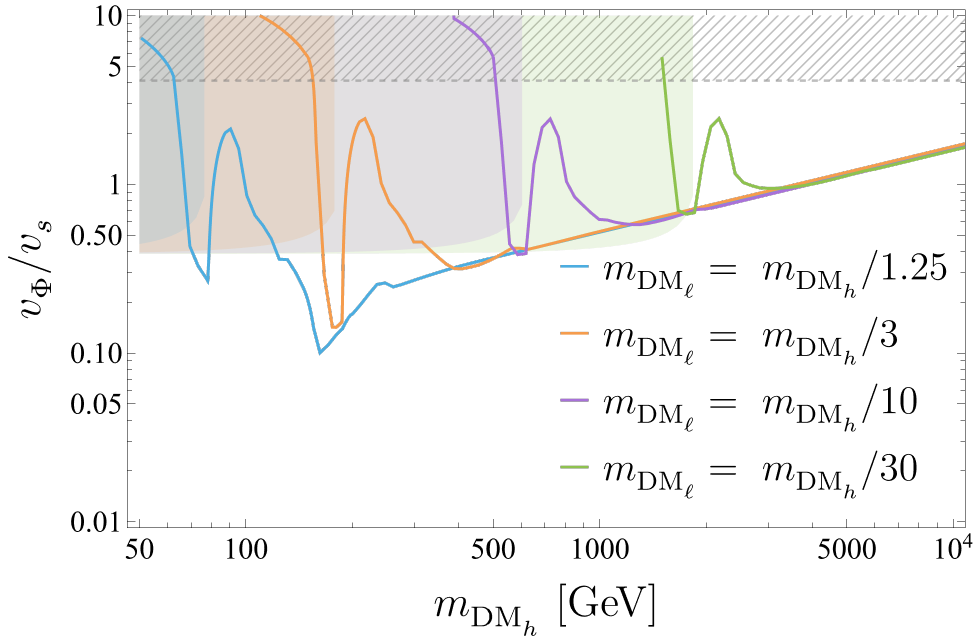}
    \caption{Heavy DM species mass vs. coupling.}
    \label{m2:fig:relic abundance2}
  \end{subfigure}
  \caption[coupling vs. DM mass]{Parameter space consistent with the observed DM relic abundance ($\Omega_{\text{DM}} h^2 = 0.120 \pm 0.001$) and the constraints considered in this work. The parameters are fixed to $\sin \theta = 0.1, ~ m_{h_2}^{} = 300~\text{GeV},~ m_{Z'}^{} = 200~\text{GeV}$ and $\sin \epsilon = 10^{-4}$.}
  \label{m2:fig:relic abundance}
\end{figure}
\begin{figure}[t]
  \centering
  \includegraphics[width=\textwidth]{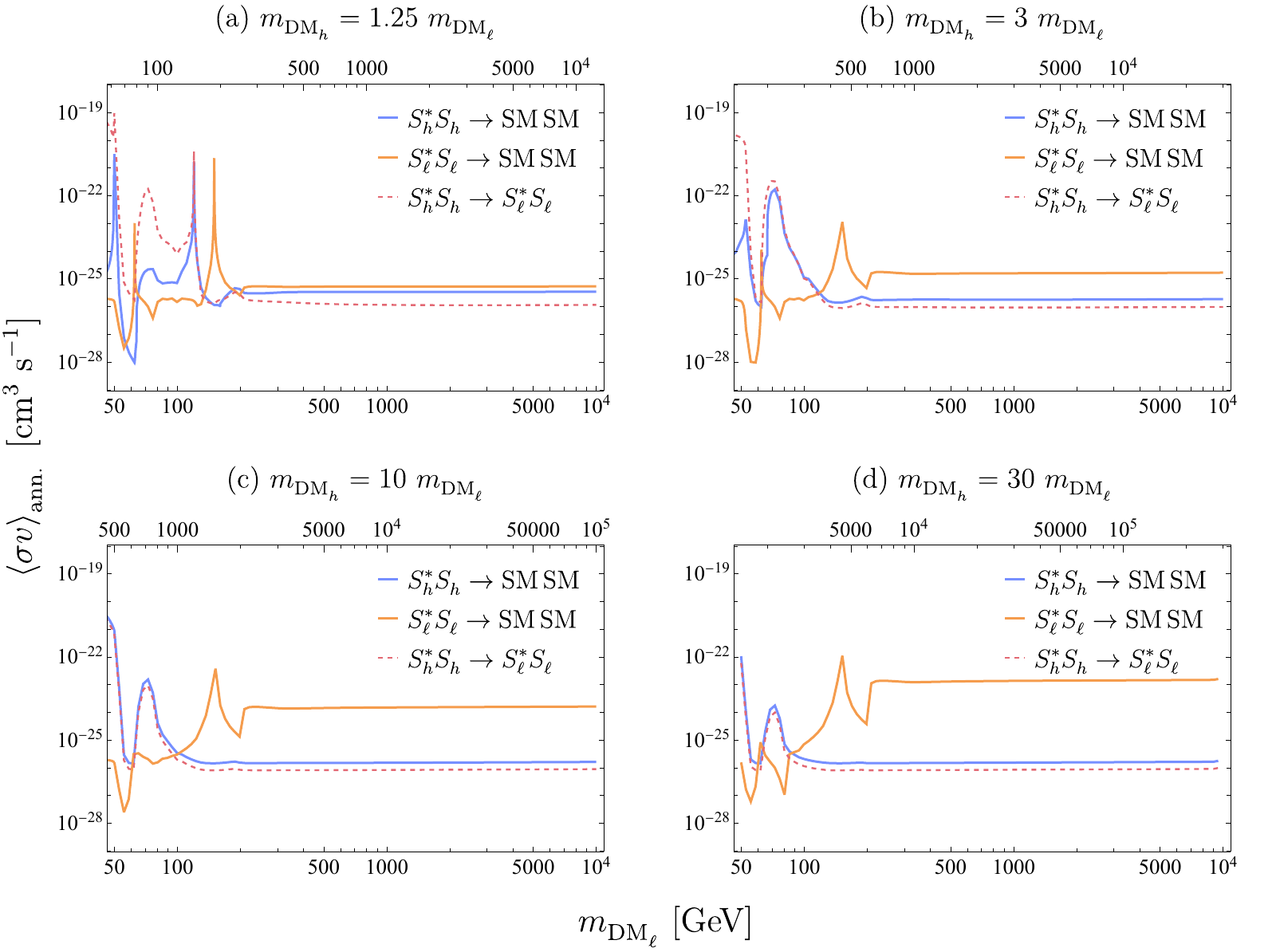}
  \caption[Annihilation vs. DM mass]{
    Annihilation cross sections of the two DM species into SM particles (solid blue: heavy, solid orange: light), along with the DM conversion cross section (dashed red) at present, corresponding to the parameter-space slices shown in \cref{m2:fig:relic abundance}. The top axis is rescaled as $m_{\text{DM}_h}^{} = \gamma m_{\text{DM}_\ell}^{}$. The parameters are fixed to $\sin\theta = 0.1$, $m_{h_2}^{} = 300~\text{GeV}$, $m_{Z'}^{} = 200~\text{GeV}$, and $\sin\epsilon = 10^{-4}$.
  }
  \label{m2:fig:csvsBoost}
\end{figure}
\begin{figure}[t]
  \centering
  \includegraphics[width=\textwidth]{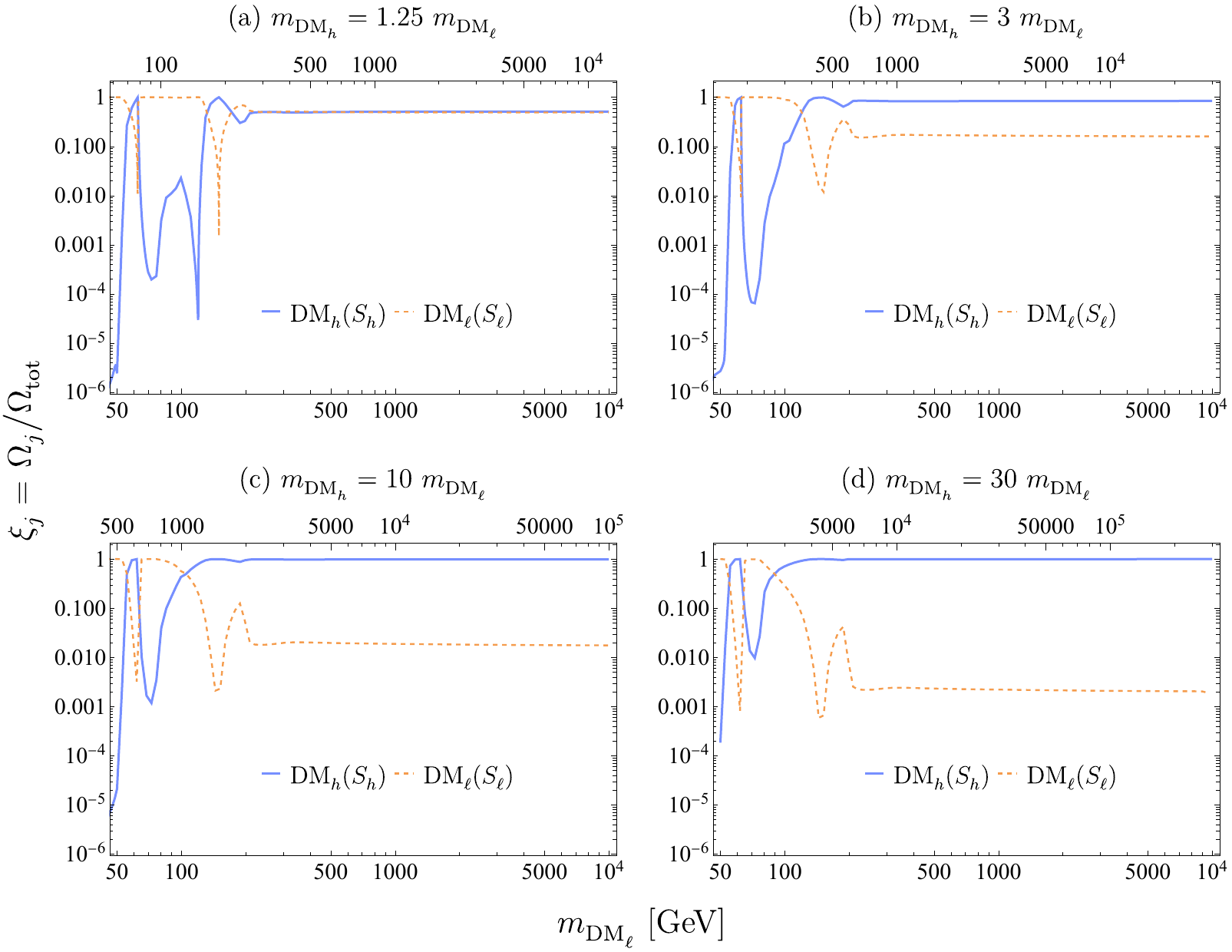}
  \caption[Fractional relic abundances vs. DM mass]{
    Fractional relic abundances of the heavier (solid blue) and lighter (dashed orange) DM species, respectively at present, corresponding to the parameter-space slices shown in \cref{m2:fig:relic abundance}. The parameters are fixed to $\sin\theta = 0.1$, $m_{h_2}^{} = 300~\text{GeV}$, $m_{Z'}^{} = 200~\text{GeV}$, and $\sin\epsilon = 10^{-4}$. The index $j = h, ~\ell$ labels the two DM species.
  }
  \label{m2:fig:fracDM}
\end{figure}

In \cref{m2:fig:relic abundance}, we present representative slices of the parameter space that reproduce the observed DM relic abundance in the plane of the light DM mass (\cref{m2:fig:relic abundance1}), heavy DM mass (\cref{m2:fig:relic abundance2}) and the coupling ratio $v_\Phi^{}/v_s^{}$ for $\gamma = 1.25, \, 3,\, 10,\, 30$. We fix the parameters in \cref{m2:eq:free-parameters} to be $\sin \theta = 0.1, ~ m_{h_2}^{} = 300~\text{GeV},~ m_{Z'}^{} = 200~\text{GeV}$ and $\sin \epsilon = 10^{-4}$. The gray hatched region is excluded by PU bounds on the scalar quartic and dark gauge couplings. In addition, the red region in \cref{m2:fig:relic abundance1} and the coloured regions in \cref{m2:fig:relic abundance2} (matching the curves) are excluded by constraints from the Higgs invisible decay. Moreover, in the nearly degenerate limit $\gamma = 1.25$, the curve has been normalized by dividing by $2^{1/4}$.\footnote{In this limit, the total relic abundance satisfies $\Omega_{\text{tot}}^{} \simeq 2\,\Omega_{i(=h,\ell)}^{}$. In the relevant region, $\Omega h^2 \propto 1/\expval{\sigma v}_\text{ann.}^{}$ and $\expval{\sigma v}_\text{ann.}^{} \propto (v_\Phi^{} / v_s^{})^4$. Consequently, relative to a single-component estimate, the coupling is shifted by a factor $2^{1/4}$. This rescaling is applied only to remove this trivial near-degeneracy effect and to facilitate comparison of the non-trivial branch structure.} As discussed earlier, the degenerate-mass limit leads to an enhanced symmetry, and thus this blue curve resembles the model of \cite{Abe:2022mlc}. Therefore, our model admits a broad parameter region consistent with the relic abundance, with its structure controlled by the boost factor $\gamma$.

The allowed band between these constraints shows a characteristic structure with dips near scalar resonances and a shift to larger couplings for larger boost factors $\gamma$. In \cref{m2:fig:csvsBoost}, we show the annihilation cross sections for the heavy DM species (solid blue) and light DM species (solid orange) to SM particles, as well as the conversion of DM species (dashed red), for the parameter-space slices presented in \cref{m2:fig:relic abundance}. It is evident that the conversion process is non-negligible throughout these slices, and it is more dominant for smaller boost factors. For larger boost factors, the light-species annihilation channel becomes comparatively more important while the conversion rate is reduced yet still relevant.

In \cref{m2:fig:fracDM}, we show the fractional relic abundances of both the DM species. The heavier and lighter DM species are represented by the solid blue and dashed orange lines, respectively. For smaller boost factors, the conversion process efficiently transfers abundance from the heavier DM species to the lighter one once the heavy component starts to depart from equilibrium, and the light-species yield can therefore be temporarily replenished.

Therefore, our model naturally predicts a scenario discussed in \cite{Agashe:2014yua} where the heavier DM species can be the dominant component of the relic abundance, while the lighter species can be subdominant yet still highly boosted, thus providing a viable BDM scenario with a large boost factor $\gamma$.

\subsection{Indirect Detection Prospects}
\label{m2:sec:indirect_detection}

\begin{figure}[t]
  \centering






  \includegraphics[width=5.2cm]{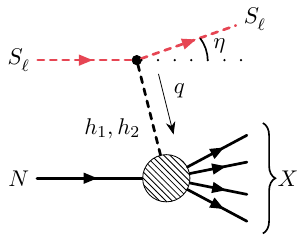}
  \caption[DM DIS]{Deep inelastic scattering of boosted dark matter $S_\ell^{}$ off a nucleon $N$, mediated by Higgs bosons $h_1^{}$ and $h_2^{}$. The process results in final-state hadronic shower $X$.}
  \label{m2:fig:DM-DIS}
\end{figure}

\begin{figure}[t]
  \centering
  \includegraphics[width=0.7\textwidth]{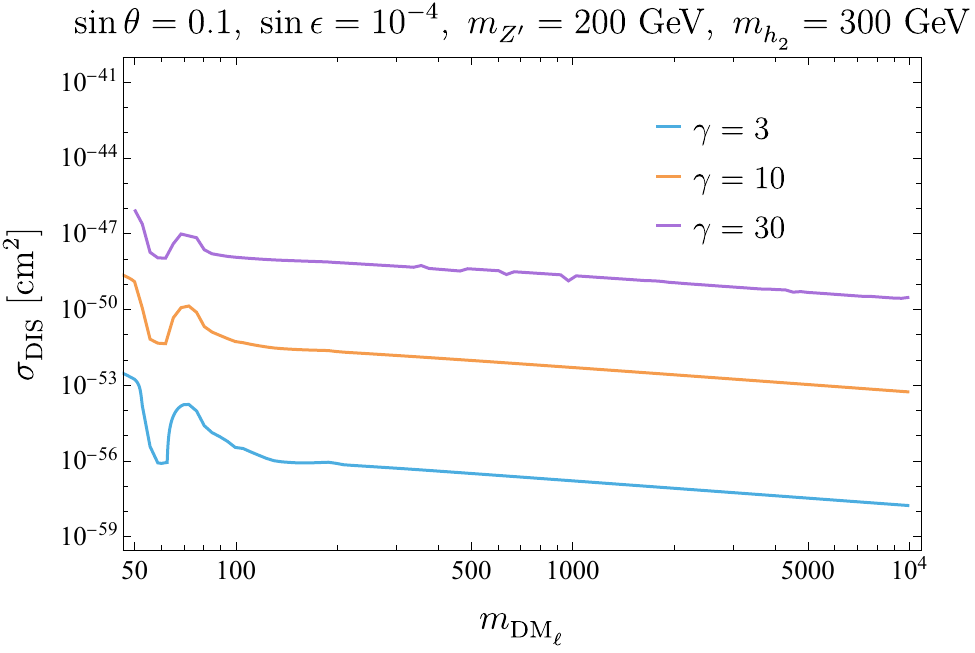}
  \caption[DIS vs. boost]{DM--Nucleon DIS cross section numerical results.}
  \label{m2:fig:dis_scat}
\end{figure}

\paragraph{Boosted Dark Matter (BDM).} The conversion process in \cref{m2:eq:conversion_processes} (see also \cref{m2:fig:DM-conversion}) can, in principle, lead to a flux of relativistic $S_\ell^{}$ particles in the present universe that can scatter off nucleons in large-volume neutrino detectors and direct-detection experiments. Assuming the initial $S_h^{}$ particles are non-relativistic, the energy of the DM particle in the final state can be kinematically determined as
\begin{equation}
  E_{\text{BDM}}^{} \equiv \gamma\, m_{\text{DM}_\ell^{}}^{} = m_{\text{DM}_h^{}}^{}.
\end{equation}
Since, these incoming $S_\ell^{}$ particles are relativistic, and their energies are well beyond the mass of the neucleons, the scattering process can be treated as deep inelastic scattering (DIS) of DM off the nucleons. The relevant Feynman diagram for this process is shown in \cref{m2:fig:DM-DIS}. Furthermore, we can define some quantites that are commonly used in the calcualtion of DIS processes \cite{HoefkenZink:2024hor}, such as
\begin{itemize}
  \item Momentum transfer: $Q^2 \equiv - q^2$.
  \item Inelasticity: $y = \flatfrac{\qty(W^2 - m_N^{2})}{\qty(2 m_N^{} E_{\text{DM}_\ell^{}} (1-x))}$.
  \item Bjorken scaling variable: $x = \flatfrac{Q^2}{\qty(Q^2 + W^2 - m_N^{2})}$.
\end{itemize}
Here, $m_N^{}$ is the mass for the nucleon and $W$ is the invariant mass of the hadronic system. We can also find the relation between the energy of the scattered BDM ($E_{\text{DM}_\ell^{}}'$) and its scattering angle ($\eta$) in terms of the above variables, i.e.,
\begin{align}
  E_{\text{DM}_\ell^{}}' & = \qty(1 - y) E_{\text{DM}_\ell^{}}^{},         \label{m2:eq:c21}           \\
  \cos \eta              & = 1 - \frac{m_N^{} xy}{\qty(1-y)E_{\text{DM}_\ell^{}}^{}}.\label{m2:eq:c22}
\end{align}
Therefore, we will calculate the DIS cross-section in the $x$-$y$ plane where the PDFs are only dependent on the Bjorken scaling variable.
The squared amplitude for the partonic process $S_\ell^{}\,q \to S_\ell^{}\,q$ can be written as
\begin{equation}
  \label{m2:eq:spin agv amplitude}
  \frac{1}{2} \sum\limits_\text{spin} \abs{i\mathcal{M}}^2
  =
  \qty(\cfrac{v_\Phi^{}}{v_s^{}})^2
  \frac{\sin^2 2\theta ~\qty(m_{h_1}^2 - m_{h_2}^2)^2}{4v_\Phi^4}
  \frac{    m_q^{2}\qty(4 m_q^2 + Q^2)  Q^4}{\qty( m_{h_1}^2 + Q^2)^2 \qty(m_{h_2}^2 + Q^2)^2},
\end{equation}
where $m_q^{}$ denotes the mass of quarks. Therefore, the final expression for the DIS differential cross section becomes
\begin{align}
  \frac{\dd{\sigma_{S_\ell^{} N}^{}}}{\dd{x}\dd{y}}
  =
  \cfrac{
    y
    \sin^2 2\theta
    \qty(1 - \cfrac{m_{h_1}^{2}}{m_{h_2}^{2}})^2
    \qty(\cfrac{v_\Phi^{}}{v_s^{}})^2
  }{64 \pi \qty(1 - \cfrac{m_{\text{DM}_\ell^{}}^{2} }{E_{\text{DM}_\ell^{}}^2}) \,v_\Phi^4 m_{h_1}^{4}}
  \sum_q
  \frac{ m_q^2 ~Q^4 ~f_q^{}(x, Q^2)}{\qty(1 + \cfrac{Q^2}{m_{h_1}^{2}})^2 \qty(1 + \cfrac{Q^2}{m_{h_2}^{2}})^2 },
  \label{m2:eq:DIS_final_integrand}
\end{align}
where, $f_q(x, Q^2)$ is the PDF for quark flavor $q$ in nucleon $N$ at momentum fraction $x$ and scale $Q^2$.
We can calculate the total DIS cross section numerically by integrating \cref{m2:eq:DIS_final_integrand} over the kinematically allowed ranges of $x$ and $y$, i.e.,
\begin{align}
  \frac{Q^2_{\min}}{Q_{\min}^2 + W_{\max}^2 - m_N^2} < {}                              & x < \frac{Q^2_{\max}}{Q_{\max}^2 + W_{\min}^2 - m_N^2}, \label{m2:eq:bjorken-x-range}                                                                                   \\
  \frac{W^2_{\min} - m_N^{2}}{2 m_N^{} \qty(1-x) \gamma m_{\text{DM}_\ell^{}}^{}}  <{} & y  < \frac{1 -\cfrac{1}{\gamma^2}}{1 + \cfrac{x m_N^{}}{2 \gamma m_{\text{DM}_\ell^{}}^{}} + \cfrac{m_{\text{DM}_\ell^{}}^{}}{2 x m_N^{} \gamma}},\label{m2:eq:y-range}
\end{align}
using \texttt{LHAPDF v6} \cite{Buckley:2014ana} with the \texttt{cteq6l1} PDF set. In \cref{m2:fig:dis_scat}, we present the numerical result of DIS cross section for $\gamma = 3, ~10, ~30$ considering the typical values of the parameters, i.e., $\sin \theta = 0.1, ~\sin \epsilon = 10^{-4}, ~m_{Z'} = 200~\text{GeV}$ and $m_{h_2}^{} = 300~\text{GeV}$. We find the cross section to be of $\sim \order{10^{-50}}~\text{cm}^2$ for the BDM-motivated region of the model.

The diffuse halo flux is too small to compensate for the suppressed DIS cross section, so we consider compact DM overdensities around black holes as sources of boosted $S_\ell^{}$. In particular, we study the conversion process in \cref{m2:eq:conversion_processes} ($S_h^* S_h^{} \to S_\ell^* S_\ell^{}$) inside a spike sourced by the heavy DM species $S_h^{}$. The spike profile for a Supermassive Black Hole (SMBH) was first studied in \cite{Gondolo:1999ef}, and the same framework has also been applied to Intermediate Mass Black Holes (IMBHs) \cite{Aschersleben:2024xsb,Vecchi:2025rzi}.

Following the standard black-hole-spike formalism of \cite{Gondolo:1999ef,Aschersleben:2024xsb,Vecchi:2025rzi}, and assuming an NFW host halo, we adopt the spike profile
\begin{equation}
  \rho_\text{sp}^{}(r) = \rho_{\text{halo}}^{}(r_\text{sp}^{}) \qty(\cfrac{r}{r_\text{sp}^{}})^{-\gamma_\text{sp}^{}},
  \label{m2:eq:spike_profile}
\end{equation}
where, $r_\text{sp}^{} \sim 0.2 r_\text{h}^{}$ is the spike radius defined in terms of the radius of influence $r_\text{h}^{}$ of the black hole, and $\rho_{\text{halo}}^{}(r_\text{sp}^{})$ is the halo density at the spike radius. The spike slope $\gamma_\text{sp}^{}$ is related to the inner slope $\gamma_\text{halo}^{}$ of the host halo with $\gamma_\text{sp}^{} = (9 - 2\gamma_\text{halo}^{})/(4 - \gamma_\text{halo}^{})$, so that $\gamma_\text{sp}^{} = 7/3$ for an NFW cusp. Annihilation of the heavy DM species $S_h^{}$ saturates the spike at small radii and is given by
\begin{equation}
  \label{m2:eq:saturation_radius}
  r_\text{sat}^{} = r_\text{sp}^{} \qty{
    \qty(\cfrac{m_{\text{DM}_h^{}}^{}}{\ev{\sigma v}_{\text{ann}}^{}})
    \qty(\cfrac{1}{t_f^{} - t_i^{}})
    \qty(\cfrac{1}{\rho_{\text{halo}}^{}(r_\text{sp}^{})})
  }^{-1/\gamma_\text{sp}^{}},
\end{equation}
with $t_f^{}$ being the age of the universe and $t_i^{}$ being the formation time of the black hole \cite{Gondolo:1999ef}.
Below $r_\text{sat}^{}$ we assume the standard annihilation-softened weak cusp. For a black hole at distance $D$, the source can be treated as point-like and the boosted flux is
\begin{equation}
  \Phi_\text{BDM}^{} =
  \frac{\ev{\sigma v}_{\text{conv.}}^{}}{2 m_{\text{DM}_h^{}}^{2} D^2}
  \int \dd E ~ \dv{N_{\text{DM}_\ell^{}}}{E}
  \int_{2r_{\text{schw}}^{}}^{r_\text{sp}^{}} \dd r ~r^2 \rho^2(r).
\end{equation}
For the monochromatic final state in \cref{m2:eq:conversion_processes}, one has $\dv*{N_{\text{DM}_\ell^{}}}{E} = \delta(E - m_{\text{DM}_h^{}})$. Using the standard spike-plus-weak-cusp profile and taking $r_\text{cut}^{} = r_\text{sat}^{} \gg 2r_{\text{schw}}^{}$, the flux reduces to
\begin{equation}
  \Phi_\text{BDM}^{} =
  \qty(\frac{\ev{\sigma v}_{\text{conv.}}^{}}{m_{\text{DM}_h^{}}^{2}})
  \frac{1}{D^2}
  \cfrac{\rho_{\text{halo}}^2(r_\text{sp}^{})}{4\gamma_\text{sp}^{} - 6}
  \,r_\text{sp}^3
  \qty{\qty(2\gamma_\text{sp}^{} - 1)
    \qty(\cfrac{r_\text{sat}^{}}{r_\text{sp}^{}})^{3 - 2\gamma_\text{sp}^{}} - 2}.
\end{equation}
Since $r_\text{sat}^{} \propto \qty(m_{\text{DM}_h^{}}^{}/\ev{\sigma v}_{\text{ann}}^{})^{-1/\gamma_\text{sp}^{}}$, the total scaling becomes
\begin{equation}
  \Phi_\text{BDM}^{} \propto \ev{\sigma v}_{\text{conv.}}^{} \times \ev{\sigma v}_{\text{ann}}^{-5/7} \times m_{\text{DM}_h^{}}^{-9/7},
\end{equation}
so a smaller annihilation rate delays saturation and enhances the boosted source. In the heavy-mass regime, in this model we can naturally achieve $\ev{\sigma v}_{\text{ann}}^{} \simeq \ev{\sigma v}_{\text{conv.}}^{}$ (see, \cref{m2:fig:csvsBoost}), this further simplifies to
\begin{equation}
  \Phi_\text{BDM}^{} \propto \ev{\sigma v}_{\text{conv.}}^{2/7} \times m_{\text{DM}_h^{}}^{-9/7}.
\end{equation}

\begin{table}[t]
  \centering
  \begin{tabular}{lcc}
    \toprule
    \textbf{Parameter}                          & \textbf{IMBH}                                           & \textbf{SMBH}                      \\
    \midrule
    Mass ($M_{\text{BH}}^{}$)                   & $8.2 \times 10^3 \, M_{\odot}^{}$ \cite{Vecchi:2025rzi} & $4.15 \times 10^6 \, M_{\odot}^{}$ \\
    Distance ($D$)                              & $5.43$ kpc                                              & $8.0$ kpc                          \\
    Schwarzschild Radius ($r_{\text{schw}}^{}$) & $\sim 7.84 \times 10^{-10}$ pc                          & $\sim 3.97 \times 10^{-7}$ pc      \\
    Radius of Influence ($r_\text{h}^{}$)       & $\sim 4.76$ pc                                          & $\sim 107.13$ pc                   \\
    Spike Radius ($r_\text{sp}^{}$)             & $\sim 0.95$ pc                                          & $\sim 21.43$ pc                    \\
    \bottomrule
  \end{tabular}
  \caption{Astrophysical parameters used for the black hole benchmarks.}
  \label{m2:tab:bh_candidates}
\end{table}

To estimate the observable flux, we consider two benchmark black holes: a possible IMBH in $\omega$-Centauri and the SMBH Sgr A* at the Galactic Center (see \cref{m2:tab:bh_candidates}). The former provides a relatively clean target, while the latter maximizes the possible luminosity.

\begin{figure}[t]
  \centering
  \begin{subfigure}{0.49\textwidth}
    \includegraphics[width=\linewidth]{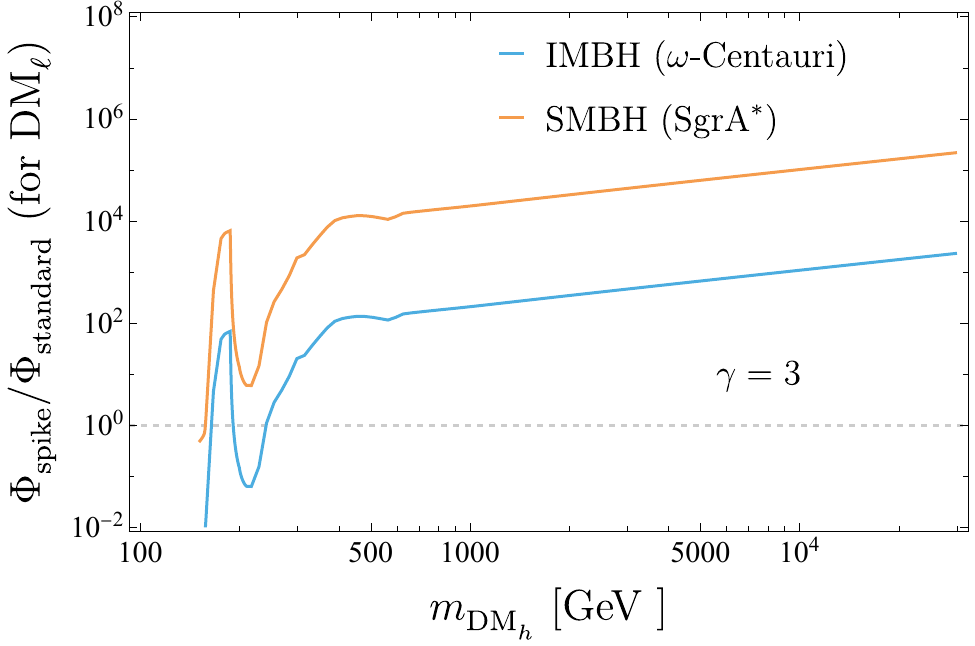}
    \caption{Boost factor $\gamma = 3$.}
  \end{subfigure}\hfill
  \begin{subfigure}{0.49\textwidth}
    \includegraphics[width=\linewidth]{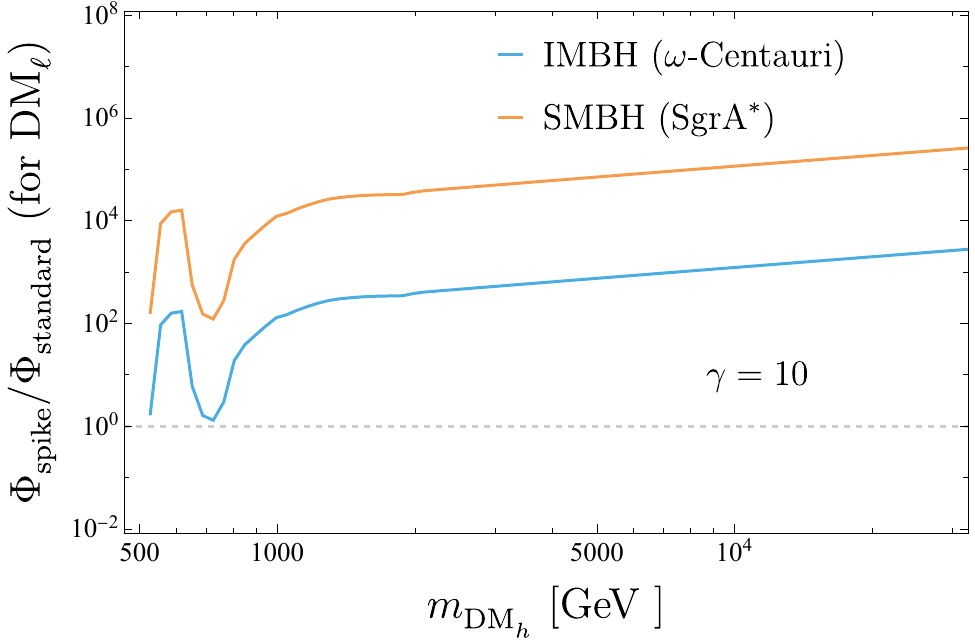}
    \caption{Boost factor $\gamma = 10$.}
  \end{subfigure}\\[0.8em]
  \begin{subfigure}{0.49\textwidth}
    \includegraphics[width=\linewidth]{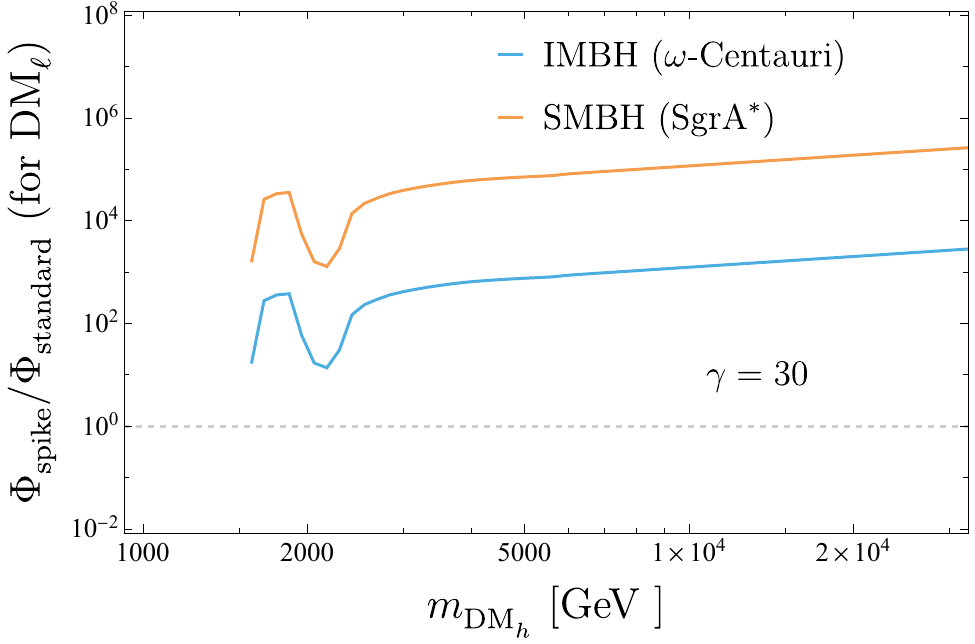}
    \caption{Boost factor $\gamma = 30$.}
  \end{subfigure}
  \caption{Calculated flux ratio of BDM from black hole spike with standard halo as a function of the heavier dark matter mass. The blue curve denotes the IMBH spike, the orange curve the Sgr A* spike.}
  \label{m2:fig:mass_BDMFlux}
\end{figure}

In \cref{m2:fig:mass_BDMFlux}, we present the flux from the SMBH spike (orange) and IMBH spike (blue) for $\gamma = 3, ~10 ,~30$. The flux is enhanced by about $10^5$ and $10^3$, respectively, relative to the standard diffuse halo contribution \cite{Agashe:2014yua}. In conventional WIMP scenarios such an enhancement would often be tightly constrained by gamma-ray observations \cite{Fields:2014pia}. In our pNGB setup, however, the dominant conversion process populates the invisible dark-sector state $S_2^{}$, allowing a sizable BDM flux without an accompanying visible gamma-ray signal. The overall normalization is still subject to the usual astrophysical uncertainties associated with the formation and survival of the spike.

With the boosted flux, assuming an exposure time of 10 years and large-volume neutrino detectors such as KM3NeT, the expected number of DIS events for BDM with mass $\gtrsim \order{100} ~\text{GeV}$ is $\order{10^{-6}}$ for an IMBH source and $\order{10^{-4}}$ for an SMBH source, which remains well below the observable level. Thus, even after accounting for black-hole spike enhancement, the event rate is negligibly small, motivating the need for larger detector volumes or longer exposure times. This suppression originates from the quark-mass-squared dependence of the DIS cross section as in \cref{m2:eq:DIS_final_integrand}, together with the dominance of the light-quark PDF $f_q(x,Q^2)$.

\paragraph{Other Phenomenological Impacts.}
The two DM species in our model can also annihilate to SM particles via the Higgs portal and are therefore in principle subject to indirect detection constraints from gamma-ray and cosmic-ray searches. In the present setup, however, the usual single-component limits are not directly applicable, because the observable signal is shared between two DM species and is weighted by their present-day relic abundances. In addition, the conversion process in \cref{m2:eq:conversion_processes} ($S_h^* S_h^{} \to S_\ell^* S_\ell^{}$) does not itself produce visible SM particles, and when annihilation proceeds through intermediate dark-sector states the resulting photon spectrum can be softer than the standard prompt $\tau\overline{\tau}$ or $q\overline{q}$ channels constrained by Fermi-LAT \cite{Fermi-LAT:2015att}. For these reasons, indirect detection bounds on the parameter space can be weaker than in the usual one-component Higgs-portal case. A quantitative assessment would require a dedicated recast and is left for future work.


\section{Conclusion}
\label{m2:sec:conclusion}

We have studied a two-component pNGB DM model based on a complex scalar field charged under a dark $\mathrm{U}(1)_V$ gauge symmetry and a global $\mathrm{SU}(3)_g$ symmetry which is softly broken to $\mathrm{U}(1)_3\times \mathrm{U}(1)_8$ symmetry. This residual symmetry automatically stabilizes both states and permits the conversion process in \cref{m2:eq:conversion_processes} ($S_h^* S_h^{} \to S_\ell^* S_\ell^{}$). The central motivation was to realize a BDM framework within the pNGB paradigm while keeping the heavier component sufficiently abundant today. In the present setup, this scenario is naturally realized. Compared with more generic multi-component portal models, where the relic fractions are often controlled by several independent couplings and viable regions are frequently tied to hierarchical portal choices \cite{Bhattacharya:2013twoComp,Bian:2014multiHiggs,Nagao:2024hit}, the heavier-component abundance in our model follows more directly from the symmetry structure and coupled freeze-out dynamics.

After imposing the constraints considered in this work, namely perturbative unitarity and Higgs invisible decay together with the observed relic abundance, we found parameter regions compatible with these requirements. The relic abundance was computed using \texttt{micrOMEGAs}, showing that conversion remains non-negligible across the representative parameter-space slices studied here.

Phenomenologically, the DM conversion process provides a concrete source of boosted $S_2^{}$ particles. We evaluated their DIS cross section on nucleons and the corresponding flux from compact DM overdensities around black holes. Although the diffuse halo contribution is too small to be observable, black-hole spikes can enhance the flux substantially, especially for Sgr A$^*$, subject to the usual astrophysical uncertainties associated with spike formation and survival. Even so, the expected event rate in a large-volume neutrino detector such as KM3NeT remains low in the benchmark configurations considered here, indicating that near-term detection is challenging. Nevertheless, the model provides a concrete proof of principle that a pNGB dark sector can naturally realize the ingredients needed for BDM and offers a motivated alternative to multi-component setups in which the required heavy or light hierarchy must be arranged more by hand.

\acknowledgments

This work was supported in part by JSPS Grant-in-Aid for Scientific Research KAKENHI Grants No. JP22K03620 (K.T.), 25H02179 (T.T.) and 25K07279 (T.T.). This work was supported by the Sasakawa Scientific Research Grant from The Japan Science Society.

\appendix
\crefalias{section}{appendix}


\section{Vacuum analysis}
\label{m2:sec:vacuum_analysis}

To analyse the vacuum alignment, we will focus on the complex scalar field $S$-dependent part of the potential. We then parametrize the VEV in the most general form as
\begin{equation}
  \ev{S} = \frac{v_s^{}}{\sqrt{2}} \mqty[
    \sin \alpha \cos \beta \\
    \sin \alpha \sin \beta \\
    \cos \alpha
  ],
\end{equation}
with $v_s^{} \ge 0$ and $\alpha,\beta \in [0,\pi/2]$. In the basis of \cref{m2:eq:soft-breaking-diagonal}, phases do not change the potential, hence we can safely remove them from the analysis. Substituting the above ansatz into \cref{m2:eq:8}, the vacuum energy becomes
\begin{align}
  \label{m2:eq:vacuum_potential}
  V(\alpha,\beta,v_s^{}) \supset {} &
  \frac{\mu_S^2}{2} v_s^2
  + \frac{\lambda_S}{8} v_s^4
  + \frac{m_3^2}{2} v_s^2 \cos 2\beta \sin^2\alpha
  - \frac{m_8^2}{12} v_s^2 \qty(1 + 3 \cos 2\alpha),
\end{align}
and the stationarity conditions are
\begin{align}
  \pdv{V}{\alpha}
  ={} & v_s^2  \,\sin 2\alpha \,\qty(m_8^2 + m_3^2 \cos 2\beta) = 0, \label{m2:eq:stationary_alpha} \\
  \pdv{V}{\beta}
  ={} & -m_3^2 v_s^2 \,\sin^2\alpha \,\sin 2\beta = 0, \label{m2:eq:stationary_beta}                \\
  \pdv{V}{v_s^{}}
  ={} & v_s^{} \qty{
                 \mu_S^2 + \frac{\lambda_S}{2} v_s^2
                 - \frac{m_8^2}{6} \qty(1 + 3 \cos 2\alpha)
                 + m_3^2 \cos 2\beta \sin^2\alpha
               } = 0. \label{m2:eq:stationary_vs}
\end{align}

Assuming a non-trivial vacuum with $v_s^{} \neq 0$ and a generic point away from the symmetry-enhanced line $m_3^2 = 0$ or $m_3^2 = \pm m_8^2$, we find that \cref{m2:eq:stationary_beta} generates the following two branches:
\begin{itemize}
  \item \textbf{Branch A.}  From \cref{m2:eq:stationary_beta} we can have
        \begin{equation}
          \sin^2 \alpha = 0 \implies \alpha = n \pi, \qquad n \in \mathbb{Z}.
        \end{equation}
        However, in the physical domain of $\alpha \in [0, \pi/2]$, the only physical solution is $\alpha = 0 ~\forall ~\beta \in [0,\pi/2]$ which corresponds to a vacuum aligned with the third direction,
        \begin{equation}
          \ev{S} = \frac{v_s^{}}{\sqrt{2}} \mqty[0 \\ 0 \\ 1],
        \end{equation}
        For this to be as local minimum, the following two provides the non-trivial conditions for the local stability of the vacuum,
        \begin{align}
          \eval{\pdv[2]{V}{\alpha}}_{\alpha=0} & = m_8^2 + m_3^2 \cos 2\beta > 0 \implies m_8^2 > \abs{m_3^2} \quad \qty(\because \beta \in [0,\pi/2]), \\
          \eval{\pdv[2]{V}{v_s^{}}}_{\alpha=0} & =  \mu_S^{2} + \frac{3}{2} v_s^2 \lambda_S - \frac{2}{3} m_8^2 > 0 \implies
          v_s^2 > \cfrac{\cfrac{2}{3} m_8^2 - \mu_S^2}{\cfrac{3}{2} \lambda_S} > 0.
        \end{align}
        Since, $v_s^{} \ge 0$, the second condition implies $\flatfrac{2 m_8^2}{3} > \mu_S^2 > 0$, and the first condition implies $m_3^2$ is bounded in the range $- m_8^2 < m_3^2 < m_8^2$. In this branch, the physical masses for the pNGBs can be calculated as,
        \begin{equation}
          m_{\text{pNGB}_1}^2 = m_8^2 + m_3^2, \qquad
          m_{\text{pNGB}_2}^2 = m_8^2 - m_3^2.
        \end{equation}
  \item \textbf{Branch B.} From \cref{m2:eq:stationary_beta} we can have
        \begin{equation}
          \sin 2\beta = 0 \implies \beta = \frac{n \pi}{2}, \qquad n \in \mathbb{Z}.
        \end{equation}
        However, the physical domain is $\beta \in [0, \pi/2]$, hence the only solutions are $\beta = 0$ and $\beta = \pi/2$. Let us discuss them separately:
        \begin{itemize}
          \item For $\beta = 0$, \cref{m2:eq:stationary_alpha} reduces to
                \begin{equation}
                  v_s^2 \, \qty(m_8^2 + m_3^2) \sin 2\alpha = 0.
                \end{equation}
                For the generic case $m_8^2 + m_3^2 \neq 0$, the solutions are $\alpha = 0$ and $\alpha = \pi/2$. The first one coincides with branch A, while the second gives
                \begin{equation}
                  \ev{S} = \frac{v_s^{}}{\sqrt{2}} \mqty[1 \\ 0 \\ 0].
                \end{equation}
                For this to be a local minimum, it should satisfy the following non-trivial conditions, i.e.,
                \begin{align}
                  \eval{\pdv[2]{V}{\beta}}_{\alpha=\pi/2,\beta=0}
                   & = - 2 m_3^2 \,v_s^2 > 0 \implies m_3^2 < 0,                   \\
                  \eval{\pdv[2]{V}{\alpha}}_{\alpha=\pi/2,\beta=0}
                   & = - v_s^2 \,\qty(m_8^2 + m_3^2) > 0 \implies m_3^2 < - m_8^2,
                \end{align}
                and
                \begin{align}
                  \eval{\pdv[2]{V}{v_s^{}}}_{\alpha=\pi/2,\beta=0}
                  ={}      & \mu_S^{2} + \frac{3}{2} v_s^2 \lambda_S + \frac{m_8^2 }{3} + m_3^2> 0,
                  \nonumber                                                                         \\
                  \implies &
                  v_s^2 > \cfrac{- \qty(\cfrac{m_8^2}{3}  + m_3^2) - \mu_S^2}{\cfrac{3}{2} \lambda_S} > 0.
                \end{align}
                Since, $v_s^{} \ge 0$, the last condition implies $- \qty(\flatfrac{m_8^2}{3}  + m_3^2) > \mu_S^2 > 0$, and the first two conditions imply $m_3^2 < 0$ and $m_8^2 < - m_3^2$. In this branch, the physical masses for the pNGBs can be calculated as,
                \begin{equation}
                  m_{\text{pNGB}_1}^2 = - 2 m_3^2, \qquad
                  m_{\text{pNGB}_2}^2 = - \qty(m_3^2 + m_8^2).
                \end{equation}
                Since, $m_3^2 < 0$ and $m_8^2 < - m_3^2$, both pNGBs are non-tachyonic.
          \item For $\beta = \pi/2$, \cref{m2:eq:stationary_alpha} reduces to
                \begin{equation}
                  v_s^2 \, \qty(m_8^2 - m_3^2) \sin 2\alpha = 0.
                \end{equation}
                For the generic case $m_8^2 - m_3^2 \neq 0$, the solutions are $\alpha = 0$ and $\alpha = \pi/2$. The first one coincides with branch A, while the second gives
                \begin{equation}
                  \ev{S} = \frac{v_s^{}}{\sqrt{2}} \mqty[0 \\ 1 \\ 0].
                \end{equation}
                For this to be a local minimum, it should satisfy the following non-trivial conditions, i.e.,
                \begin{align}
                  \eval{\pdv[2]{V}{\beta}}_{\alpha=\pi/2,\beta=\pi/2}
                   & = 2 m_3^2 \,v_s^2 > 0 \implies m_3^2 > 0,                   \\
                  \eval{\pdv[2]{V}{\alpha}}_{\alpha=\pi/2,\beta=\pi/2}
                   & = - v_s^2 \,\qty(m_8^2 - m_3^2) > 0 \implies m_8^2 < m_3^2,
                \end{align}
                and
                \begin{align}
                  \eval{\pdv[2]{V}{v_s^{}}}_{\alpha=\pi/2,\beta=\pi/2}
                  ={}      & \mu_S^{2} + \frac{3}{2} v_s^2 \lambda_S + \frac{m_8^2 }{3} - m_3^2> 0,
                  \nonumber                                                                         \\
                  \implies &
                  v_s^2 > \cfrac{- \qty(\cfrac{m_8^2}{3}  - m_3^2) - \mu_S^2}{\cfrac{3}{2} \lambda_S} > 0.
                \end{align}
                Since, $v_s^{} \ge 0$, the last condition implies $- \qty(\flatfrac{m_8^2}{3}  - m_3^2) > \mu_S^2 > 0$, and the first two conditions imply $m_3^2 > 0$ and $m_8^2 < m_3^2$. In this branch, the physical masses for the pNGBs can be calculated as,
                \begin{equation}
                  m_{\text{pNGB}_1}^2 = 2 m_3^2, \qquad
                  m_{\text{pNGB}_2}^2 = m_3^2 - m_8^2.
                \end{equation}
                Since, $m_3^2 > 0$ and $m_8^2 < m_3^2$, both pNGBs are non-tachyonic.
        \end{itemize}
\end{itemize}

\begin{table}[t]
  \centering
  \renewcommand{\arraystretch}{1.4}
  \resizebox{\linewidth}{!}{%
    \begin{tabular}{@{}l c c l@{}}
      \toprule
      Stability conditions & Orientation                                               & Vacuum configuration & pNGB masses                                  \\
      \midrule
      $m_8^2 > \abs{m_3^2}$
                           & \multirow{2}{*}{$\alpha = 0, ~\beta \in [0,\pi/2]$}
                           & \multirow{2}{*}{$\cfrac{v_s^{}}{\sqrt{2}}\mqty[0,0,1]^T$}
                           & $m_{\text{pNGB}_1}^2 = m_8^2 + m_3^2$                                                                                           \\
      $\flatfrac{2 m_8^2}{3} > \mu_S^2 > 0$
                           &                                                           &                      & $m_{\text{pNGB}_2}^2 = m_8^2 - m_3^2$        \\
      \midrule
      $m_3^2 < 0, ~ m_8^2 < -m_3^2$
                           & \multirow{2}{*}{$\alpha = \pi/2, ~\beta = 0$}
                           & \multirow{2}{*}{$\cfrac{v_s^{}}{\sqrt{2}}\mqty[1,0,0]^T$}
                           & $m_{\text{pNGB}_1}^2 = -2 m_3^2$                                                                                                \\
      $-\qty(\flatfrac{m_8^2}{3} + m_3^2) > \mu_S^2 > 0$
                           &                                                           &                      & $m_{\text{pNGB}_2}^2 = -\qty(m_3^2 + m_8^2)$ \\
      \midrule
      $m_3^2 > 0, ~ m_8^2 < m_3^2$
                           & \multirow{2}{*}{$\alpha = \pi/2, ~\beta = \pi/2$}
                           & \multirow{2}{*}{$\cfrac{v_s^{}}{\sqrt{2}}\mqty[0,1,0]^T$}
                           & $m_{\text{pNGB}_1}^2 = 2 m_3^2$                                                                                                 \\
      $-\qty(\flatfrac{m_8^2}{3} - m_3^2) > \mu_S^2 > 0$
                           &                                                           &                      & $m_{\text{pNGB}_2}^2 = m_3^2 - m_8^2$        \\
      \bottomrule
    \end{tabular}
  }
  \caption{Classification of the locally stable vacua.}
  \label{m2:tab:vacuum-classification}
\end{table}

The above analysis is summarized in \cref{m2:tab:vacuum-classification}. Outside the listed parameter regions, at least one fluctuation mode becomes tachyonic, so the corresponding stationary point is not a local minimum. A useful feature of branch A is that the symmetry-enhanced line $m_3^2 = 0$ still lies in the stable region, where the vacuum remains aligned as $\ev{S} \propto \mqty[0,0,1]^T$ and both pNGBs stay massive, $m_{\text{pNGB}_1}^2 = m_{\text{pNGB}_2}^2 = m_8^2$. By contrast, the two branch-B vacua require $m_3^2 \neq 0$ for strict local stability, and the mode with mass-squared $2\abs{m_3^2}$ becomes massless as $m_3^2 \to 0$. For this reason, and to keep the analysis in a region that remains robust even near the enhanced-symmetry limit, we take branch A as the reference vacuum in \cref{m2:sec:residual-symmetry}.


\section{Conversion of heavier DM to lighter}
\label{m2:sec:DM-DM conversion}

Here we will show the DM--DM cross section for the processes shown in \cref{m2:fig:DM-conversion} i.e.,
\begin{equation}
  S_1^* (p) S_1^{} (k) \to \qty{h_1^{},\, h_2^{},\, Z',\, \text{contact}} \to S_2^* (p') S_2^{} (k'),
\end{equation}
with four momenta as indicated in the parentheses which leads to $p^2 = k^2 = m_{\text{DM}_h^{}}^{2}$ and $p'^2 = k'^2 = m_{\text{DM}_\ell^{}}^{2}$.
Using the couplings in \cref{m2:eq:kappas}, and defining the variables to simplify the calculations,
\begin{align}
  \Delta_1 & =     - \frac{1}{v_s^2} \Bigg\{
  \frac{m_{h_1}^4 \sin^2\theta}{\qty(s - m_{h_1}^2 + i \,m_{h_1}^{} \Gamma_{h_1}^{})} +
  \frac{m_{h_2}^4 \cos^2\theta}{\qty(s - m_{h_2}^2 + i \,m_{h_2}^{} \Gamma_{h_2}^{})}
  \nonumber                                                                           \\
           & \qquad\qquad  + 2\,\qty(m_{h_1}^2 \sin^2\theta + m_{h_2}^2 \cos^2\theta)
  \Bigg\},
  \\
  \Delta_2 & = \frac{2 g_V^2}{s  - m_{Z'}^2 + i \,m_{Z'}^{} \Gamma_{Z'}^{}},
  \\
  \Delta_3 & = \frac{s}{2} - m_{\text{DM}_h^{}}^2 - m_{\text{DM}_\ell^{}}^2.
\end{align}
For the analytic discussion below, we work away from the mediator poles and adopt the zero-width approximation, $\Gamma_{h_1}^{},\Gamma_{h_2}^{},\Gamma_{Z'}^{} \to 0$. In this limit, $\Delta_1$ and $\Delta_2$ are real, and it is convenient to write them in a near-dimensionless form,
\begin{align}
  \Delta_1
           & \simeq    - ~2 ~\qty(\cfrac{v_\Phi^{}}{v_s^{}})^2 \qty[
                                                                 \frac{\qty(\cfrac{m_{h_1}^{}}{v_\Phi^{}})^2  \qty(1 - \cfrac{m_{h_1}^{2}}{2 s}) ~\sin^2\theta }{\qty(1 - \cfrac{m_{h_1}^{2}}{s})} +
                                                                 \frac{\qty(\cfrac{m_{h_2}^{}}{v_\Phi^{}})^2  \qty(1 - \cfrac{m_{h_2}^{2}}{2 s}) ~\cos^2\theta }{\qty(1- \cfrac{m_{h_2}^{2}}{s})}
                                                               ],
  \label{m2:eq:Delta1}
  \\
  \Delta_2 & \simeq \frac{2}{v_\Phi^2} \qty(\cfrac{v_\Phi^{}}{v_s^{}})^{2}
  \frac{\qty(\cfrac{m_{Z'}^{2}}{s})}{\qty(1- \cfrac{m_{Z'}^{2}}{s})},
\end{align}
where, we have used $g_V^{} \simeq \flatfrac{m_{Z'}^{}}{v_s^{}}$ arising from the smallness of the gauge kinetic mixing angle and the Lorentz boost factor $\gamma = \flatfrac{m_{\text{DM}_h^{}}^{}}{m_{\text{DM}_\ell^{}}^{}}$. The square of the amplitude then can be calculated as
\begin{equation}
  \abs{i\mathcal{M}_{S_1^* S_1^{}\to S_2^* S_2^{}}^{}}^2
  \simeq
  \qty[\Delta_1 + \Delta_2 \qty(\Delta_3 + t)]^2.
\end{equation}
Integrating the differential cross section over the physical range of $t$, i.e.,
\begin{equation}
  t_\pm = - \Delta_3 \pm \frac{\sqrt{\lambda_i \lambda_f}}{2 s},
\end{equation}
with $\lambda_i \equiv \lambda(s, m_{\text{DM}_h^{}}^2, m_{\text{DM}_h^{}}^2)$ and $\lambda_f \equiv \lambda(s, m_{\text{DM}_\ell^{}}^2, m_{\text{DM}_\ell^{}}^2)$, where $\lambda(a,b,c)=(a - b - c)^2 - 4 bc$ is the Kallen function, the total cross section for the DM--DM conversion can be written as
\begin{equation}
  \label{m2:eq:DM-DM conversion cross section}
  \sigma_\text{conv.}^{}
  =
  \int_{t_-}^{t_+} ~\dd{t} ~\dv{\sigma_\text{conv.}^{}}{t}
  = \frac{1}{16 \pi s} \sqrt{\frac{\lambda_f}{\lambda_i}} \qty(
  \Delta_1^2
  + \frac{\Delta_2^2}{12}\cfrac{\lambda_i \lambda_f}{s^2}
  ) .
\end{equation}
For the non-relativistic initial state relevant to freeze-out, $\lambda_i = s \qty(s - 4 m_{\text{DM}_h^{}}^2)$, $\lambda_f = s \qty(s - 4 m_{\text{DM}_\ell^{}}^2)$, and $s \simeq 4 m_{\text{DM}_h^{}}^2 \simeq 4 \gamma^2 m_{\text{DM}_\ell^{}}^{2}$ with $\gamma = \flatfrac{m_{\text{DM}_h^{}}^{}}{m_{\text{DM}_\ell^{}}^{}}$. Then,
\begin{align}
  \label{m2:eq:Delta terms}
  \lambda_i \simeq 4 \gamma^4 m_{\text{DM}_\ell^{}}^4 v^2,
  \quad
  \lambda_f \simeq 16 \gamma^4 m_{\text{DM}_\ell^{}}^4 \qty(1 - \frac{1}{\gamma^2}),
  \quad
  \frac{\sqrt{\lambda_i \lambda_f}}{2 s} \simeq \gamma^2 m_{\text{DM}_\ell^{}}^2 v \sqrt{1 - \frac{1}{\gamma^2}},
\end{align}
where $v$ is the DM relative velocity. Accordingly, the thermally averaged cross section in the $s$-wave approximation becomes
\begin{equation}
  \ev{\sigma v}_\text{conv.}^{}
  \simeq
  \frac{1}{32 \pi \gamma^2 m_{\text{DM}_\ell^{}}^2}
  \sqrt{1 - \frac{1}{\gamma^2}}
  \Delta_1^2 \qty(s \simeq 4 \gamma^2 m_{\text{DM}_\ell^{}}^2)
  + \order{v^2}.
\end{equation}
This shows that the $Z'$ contribution is momentum-suppressed for the conversion process, while the scalar or contact contribution encoded in $\Delta_1$ remains unsuppressed. To understand how the conversion is affected by the boost factor $\gamma$, we study the following cases:

\begin{itemize}
  \item Nearly degenerate regime ($\gamma \gtrsim 1$). Here we have $m_{\text{DM}_h^{}}^{} \simeq m_{\text{DM}_\ell^{}}^{}$ and the final-state phase space closes as $\lambda_f \propto \qty(1 - \gamma^{-2})$, so the conversion rate vanishes at the exact threshold $\gamma = 1$. However, at fixed $m_{\text{DM}_\ell^{}}^{}$ the explicit prefactor in \cref{m2:eq:Delta terms} and the equation below behaves as $\gamma^{-2} \sqrt{1-\gamma^{-2}}$, which is maximal at $\gamma = \sqrt{3/2}$. Therefore, mildly non-degenerate low-boost configurations maximize the explicit $\gamma$-dependent kinematic factor and can give the largest conversion cross section with respect to $\gamma$ at fixed mass scale and couplings provided the DM mass is itself small, consistent with the benchmark behavior where RP1 with $\gamma = 1.25$ (see, \cref{m2:fig:csvstemp}) has a much larger $11\to22$ cross section than the larger-$\gamma$ cases. If $s \simeq 4\gamma^2 m_{\text{DM}_\ell^{}}^2$ approaches one of the scalar poles, the same rate is further enhanced by the resonance structure of $\Delta_1$, producing the spikes seen in \cref{m2:fig:csvsBoost}.
  \item Highly non-degenerate regime ($\gamma \gg 1$). In this limit, the DM species are highly non-degenerate, $m_{\text{DM}_h^{}}^{} \gg m_{\text{DM}_\ell^{}}^{},~m_{h_1}^{}, ~m_{h_2}^{}, ~ m_{Z'}^{}$. Then $\lambda_f \simeq 16 \gamma^4 m_{\text{DM}_\ell^{}}^4$, while the $Z'$ contribution is still momentum-suppressed. Hence the leading conversion rate is again dominated by $\Delta_1$, but the explicit kinematic factor falls as $\sqrt{1-\gamma^{-2}}/\gamma^2 \simeq 1/\gamma^2$. Away from resonances, this implies $\sigma_\text{conv.}^{} v_{\text{rel}} \propto \gamma^{-2} m_{\text{DM}_\ell^{}}^{-2}$ up to the residual $\gamma$ dependence inside $\Delta_1$, which explains the finite conversion cross section at large boost in \cref{m2:fig:csvsBoost}.
\end{itemize}

To summarize, at fixed $m_{\text{DM}_\ell^{}}^{}$ and away from scalar resonances, the explicit kinematic prefactor behaves as $\gamma^{-2}\sqrt{1-\gamma^{-2}}$: it vanishes in the strict degenerate limit $\gamma \to 1$, reaches a maximum at $\gamma = \sqrt{3/2} \simeq 1.225$, and then decreases for larger $\gamma$. Thus the largest conversion rate is expected for mildly non-degenerate low-boost configurations provided the DM mass is itself small. Furthermore, if the DM masses are large enough, regardless of the smallness of the boost factor, the conversion cross section hits a flat platue due to the dominance of the $m_{\text{DM}_\ell^{}}^{-2}$ term (see, \cref{m2:fig:csvsBoost}), still remaining relevant.

\bibliographystyle{JHEP}
\bibliography{references.bib}

\end{document}